\documentclass{aa}  

\usepackage{graphicx,txfonts,lscape,natbib}
\bibpunct{(}{)}{;}{a}{}{,} 

\begin{document}
   \title{An eclipsing double-line spectroscopic binary at the stellar/substellar boundary in the Upper Scorpius OB association\thanks{Based on observations made with telescopes (GTC, WHT) installed in the Spanish Observatorio del Roque de los Muchachos of the Instituto de Astrof\'isica de Canarias, in the island of La Palma}}


   \author{N. Lodieu \inst{1,2}
        \and
        R.\ Alonso \inst{1,2}
        \and
        J.\ I.\ Gonz\'alez Hern\'andez \inst{1,2}
        \and
        R.\ Sanchis-Ojeda \inst{3,4}
        \and
        N.\ Narita \inst{5,6,7}
        \and
        Y.\ Kawashima \inst{8}
        \and
        K.\ Kawauchi \inst{9}
        \and
        A.\ Su\'arez Mascare\~no \inst{1,2}
        \and
        H.\ Deeg \inst{1,2}
        \and
        J.\ Prieto Arranz \inst{1,2}
        \and
        R.\ Rebolo \inst{1,2}
        \and
        E.\ Pall\'e \inst{1,2}
        \and
        V.\ J.\ S.\ B\'ejar \inst{1,2}
        \and
        A.\ Ferragamo \inst{1,2}
        \and
        J.\ A.\ Rubi\~no-Mart\'in \inst{1,2}
        }

   \institute{Instituto de Astrof\'isica de Canarias (IAC), Calle V\'ia L\'actea s/n, E-38200 La Laguna, Tenerife, Spain \\
       \email{nlodieu@iac.es}
       \and
       Departamento de Astrof\'isica, Universidad de La Laguna (ULL), E-38206 La Laguna, Tenerife, Spain
       \and
       Department of Astronomy, University of California, Berkeley, CA 94720
       \and
       NASA Sagan Fellow
       \and
       National Astronomical Observatory of Japan, 2-21-1 Osawa, Mitaka, Tokyo 181-8588, Japan
       \and
       Astrobiology Center, National Institutes of Natural Sciences, 2-21-1 Osawa, Mitaka, Tokyo 181-8588, Japan
       \and
       National Astronomical Observatory of Japan, 2-21-1 Osawa, Mitaka, Tokyo 181-8588, Japan
       \and
       SOKENDAI (The Graduate University for Advanced Studies), Shonan Village, Hayama, Kanagawa 240-0193, Japan
       \and
       Department of Earth and Planetary Science, The University of Tokyo, 7-3-1 Bunkyo-ku, Tokyo 113-0033, Japan
       \and
       Department of Earth and Planetary Sciences, Tokyo Institute of Technology, 2-12-1 Ookayama, Meguro-ku, Tokyo 152-8551, Japan 
             }

   \date{Received \today{}; accepted (date)}

 
  \abstract
   {}
   {We aim at constraining evolutionary models at low mass and young ages by identifying interesting transiting
   system members of the nearest OB association to the Sun, Upper Scorpius, targeted by the Kepler mission.}
   {We produced light curves for M dwarf members of the USco region surveyed during
   the second campaign of the Kepler K2 mission. We identified 'by eye' a transiting system,
   USco\,J161630.68$-$251220.1 (=\,EPIC\,203710387) with a combined
   spectral type of M5.25 whose photometric, astrometric, and spectroscopic properties makes it a
   member of USco. We conducted an extensive photometric and spectroscopic follow-up of this transiting
   system with a suite of telescopes and instruments to characterise the properties of each component
   of the system.}
   {We calculated a transit duration of about 2.42 hours occuring every 2.88 days with a slight 
    difference in transit depth and phase between the two components. We estimated a mass ratio
    of 0.922$\pm$0.015 from the semi-amplitudes of the radial velocity curves for each 
    component. We derived masses of 0.091$\pm$0.005 M$_{\odot}$ and 0.084$\pm$0.004 M$_{\odot}$, 
    radii of 0.388$\pm$0.008 R$_{\odot}$ and 0.380$\pm$0.008 R$_{\odot}$, luminosities of
    $\log$(L/L$_{\odot}$)\,=\,$-$2.020$_{-0.121}^{+0.099}$ dex and $-$2.032$_{-0.121}^{+0.099}$ dex, and 
    effective temperatures of 2901$_{-172}^{+199}$\,K and 2908$_{-172}^{+199}$\,K
    for the primary and secondary, respectively.
 }
   {We present a complete photometric and radial velocity characterisation of the least massive
   double-line eclipsing binary system in the young USco association with two components close to the
   stellar/substellar limit. This system fills in a gap between the least massive eclipsing binaries
   in the low-mass and substellar regimes at young ages and represents an important addition to 
   constrain evolutionary models at young ages.
   }


  \authorrunning{Lodieu et al$.$}
  \titlerunning{An eclipsing double-line low-mass spectroscopic binary in USco}

   \maketitle
%

%
%
\section{Introduction}
\label{K2transitM5:intro}

The mass of a star determines its fate. Thus, measuring masses of stars with high precision 
is of paramount importance to understand the evolution of stars at any mass and age. The 
precise determination of the mass and radius of the component of multiple systems allows 
to constrain input parameters of evolutionary models 
\citep{dantona94,palla93,baraffe98,chabrier00c,siess00,dotter08,tognelli11,bressan12,feiden12a}
and distinguish between the various models of star formation
\citep[e.g.][]{klessen01,reipurth01,padoan02,whitworth04,bate09,stamatellos09b}.
Low-mass eclipsing binaries represent an excellent test-bed for these models because both
masses and radii can be measured with high precisions \citep{torres10a}.

The advent of ground-based transiting surveys \citep[e.g.\ HAT-Net, SuperWASP;][]{bakos02,bakos04,pollacco06}
and space mission like Corot \citep{barge06a} and Kepler \citep{borucki10,lissauer14,batalha14}
revealed a significant number of low-mass stars and brown dwarfs eclipsing solar-type
stars \citep{deleuil08,bakos09b,irwin10,siverd12,steffen12a,mamajek12a,kenworthy15a}
or giants \citep{bouchy11b,mitchell03,liu08b}.
Of those only a few brown dwarfs have accurate mass and radius measurements:
Corot-3\,b \citep{deleuil08}, KELT-1\,b \citep{siverd12}, and Kepler-39\,b \citep{bouchy11b}.
Additional mass measurements of brown dwarfs are available from microlensing events
\citep[OGLE-2013-BLG-0102L\,b;][]{jung15}, and dynamical orbits 
\citep{dupuy09a,dupuy09b,konopacky10,dupuy10,dupuy14,dupuy15b}.

The number of young low-mass eclipsing systems is even more scarce, most of them identified
in the Orion star-forming region \citep{irwin07a,cargile08,morales_calderon12,gomez_maqueo12},
25\,Ori \citep{vanEyken11} and in the 3 Myr-old cluster NGC\,2264 with Corot \citep{gillen14}.
Recently, \citet{kraus15a} announced the first spectroscopic binary member of Upper Scorpius
(UScoCTIO\,5) with masses (0.317--0.329 M$_{\odot}$) and radii (0.810-0.834 R$_{\odot}$) 
for two M4.5 dwarfs, system originally classified as a M4 by \citet{ardila00}. These are the first 
ones determined independently from evolutionary models at such low-masses in that association.
Only one brown dwarf-brown dwarf eclipsing binary has been announced as a member of the 
Orion star-forming region \citep{stassun06} with a surprising temperature reversal
\citep{stassun07b}.

Kepler is a 0.95-m satellite launched in March 2009 to look for extrasolar planets
in a large field of $\sim$100 deg$^{2}$ towards the Cygnus/Lyra/Draco constellations.
The main goal of the mission was to explore the structure and diversity of planetary 
systems through the transit method with unprecedented photometric precision only 
achievable from space. The Kepler mission was very successful \citep{borucki10}
and extended by NASA despite the loss of reaction wheels. A K2 ``Second Light'' mission
started and is sub-divided into campaigns. The second campaign included in particular
most of the known USco member \citep*{luhman12c}.

The USco region is part of the nearest OB association to the Sun, Scorpius Centaurus,
located at 145 pc from the Sun \citep{deBruijne97}. Its age lies in the 5 to 10 Myr range
\citep{song12}: earlier studies using isochrone fitting and dynamical studies derived an
age of 5$\pm$2 Myr \citep{preibisch02} in agreement with deep surveys
\citep{slesnick06,lodieu08a} but recently challenged by \citet{pecaut12} who quoted
11$\pm$2 Myr from a spectroscopic study of F stars at optical wavelengths.
The association has been targeted at multiple wavelengths, starting off in X~rays 
\citep{walter94,kunkel99,preibisch98}, but also astrometrically with Hipparcos
\citep{deBruijne97,deZeeuw99}, and more recently in the optical
\citep{preibisch01,preibisch02,ardila00,martin04,slesnick06} and in the
near--infrared \citep{lodieu06,lodieu07a,dawson11,lodieu11a,dawson12,lodieu13c}.
Tens of brown dwarfs have now been confirmed spectroscopically as USco members
\citep{martin04,slesnick06,lodieu06,slesnick08,lodieu08a,martin10a,dawson11,lodieu11a}
and the mass function determined for high-mass stars \citep{preibisch02a}
as well as low-mass stars and brown dwarfs \citep{slesnick08,lodieu11a}.
The coolest member candidates so far may be late-L or T-type dwarfs \citep{lodieu13d}.

In this paper, we describe our analysis of the K2 light curves obtained in the USco region
and present the discovery of a new eclipsing binary, USco\,J161630.68$-$251220.1 (hereafter USco16m25)
with two components at the stellar/substellar limit.
In Section \ref{K2transitM5:identification} we describe the generation of light curves to 
search for transiting systems.
In Section \ref{K2transitM5:phot_followup} we present additional transits at optical 
wavelengths to confirm the eclipsing nature of the USco16m25 system.
In Section \ref{K2transitM5:spec_followup} we present optical spectroscopy of the system
which we classify as a M5.25 dwarf as well as intermediate-resolution spectroscopy to measure 
the radial velocities and constrain the mass and radii of each component of the eclipsing system.
In Section \ref{K2transitM5:characterisation} we characterise the physical parameters of
the two components of USco16m25 (masses, radii, effective temperature, and age) and place
our results in context with other low-mass and brown dwarf members of eclipsing systems.

%
%
\section{Identification of the transiting system}
\label{K2transitM5:identification}
\subsection{Extraction of K2 light curves}
\label{K2transitM5:K2_lc}

The second K2 campaign focused on a large region of the USco association. The K2 observations 
lasted 76.19 days, starting on 26 August 2014 at UT\,$\sim$\,05h10m and ending on 10 November 
2014 at UT\,$\sim$\,09h44m. One photometric point was obtained every 30\,min, i.e.\ with the
low cadence of K2\@.

\begin{figure}
  \centering
  \includegraphics[width=\linewidth, angle=0]{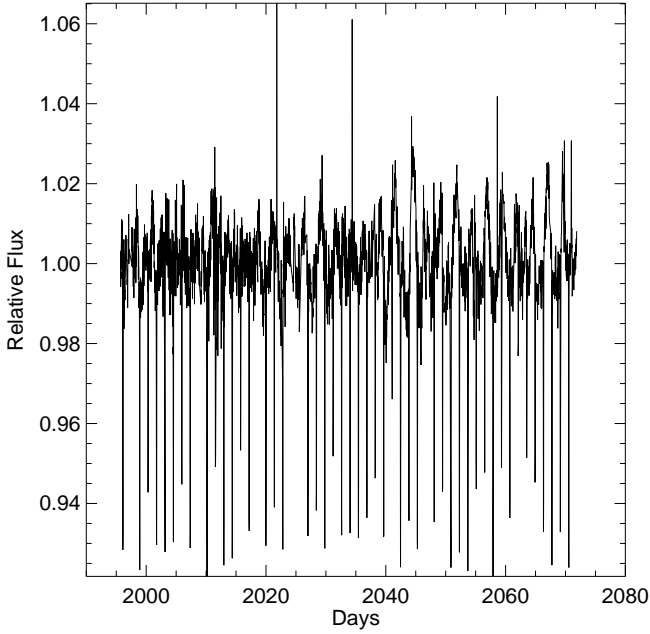}
   \caption{Kepler light-curve for USc16m25 over the full
$\sim$76 days of the K2 campaign two in Upper Scorpius.
   }
   \label{fig_K2transitM5:K2_transit}
\end{figure}

We generated the photometric light curves from the K2 observations for a large number of USco
members proposed by independent groups using our own tools.
We retrieved the images obtained with the Kepler telescope via MAST public archive on 2015 March 20\@. 
As described in detail in \citet{sanchis_ojeda15a}, we first obtain the centroids of the images and use 
them to identify images taken while the telescope was re-oriented, also known as thruster events 
\citep{howell14,vanderburg14}. After removing those images, we sliced the data in 10 temporal 
segments, chosen to have a length of approximately one week and to start (end) right after (before) a 
thurster event. We defined an aperture at each segment with all the pixels that had 4\% more counts 
than the mean background on at least 50\% of the images of that segment. After subtracting the background,
we generated a raw photometric light curve with a simple sum of the counts of each pixel. We removed
systematic effects and long term trends using 4$^{th}$ order polynomials that have the centroid displacement 
and the time as variables. For more details on how this full process work, we refer the reader to 
\citet{sanchis_ojeda15a}. Scrolling through the light curves by eye, two targets caught our 
attention: HD\,144548 \citep{alonso15a} and EPIC\,203710387 (=USco16m25), which is 
the focus of this paper. Its full light-curve is displayed in Fig.\ \ref{fig_K2transitM5:K2_transit}.

However, it is worth noting that studying stellar activity using the Kepler telescope is a 
complicated task. The Kepler photometric pipeline did not incorporate a good description of long 
term systematics until a few years after the start of the mission \citep{smith12a,stumpe12}
Our algorithm has preserved all the original long term variations found in the raw photometry, 
but these could be a combination of astrophysical and unquantified systematic variations. However, 
the lack of a prevalence of periodic variations at a period of 2.6 days among the rest of the 
K2 targets in field 2 makes us confident that our signal is mostly dominated by stellar activity. 

\subsection{Preparation of K2 light curve}
\label{K2transitM5:K2_lc_phase}

\begin{figure}
  \centering
  \includegraphics[width=\linewidth, angle=0]{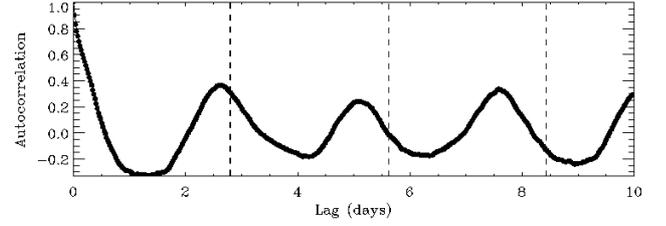}
   \caption{Autocorrelation of the K2 light curve of USco16m25\@. The vertical dashed lines 
represent multiples of the period of the system.
   }
   \label{fig_K2transitM5:K2lc_autocorrelation}
\end{figure}

\begin{figure*}
  \centering
  \includegraphics[width=0.49\linewidth, angle=0]{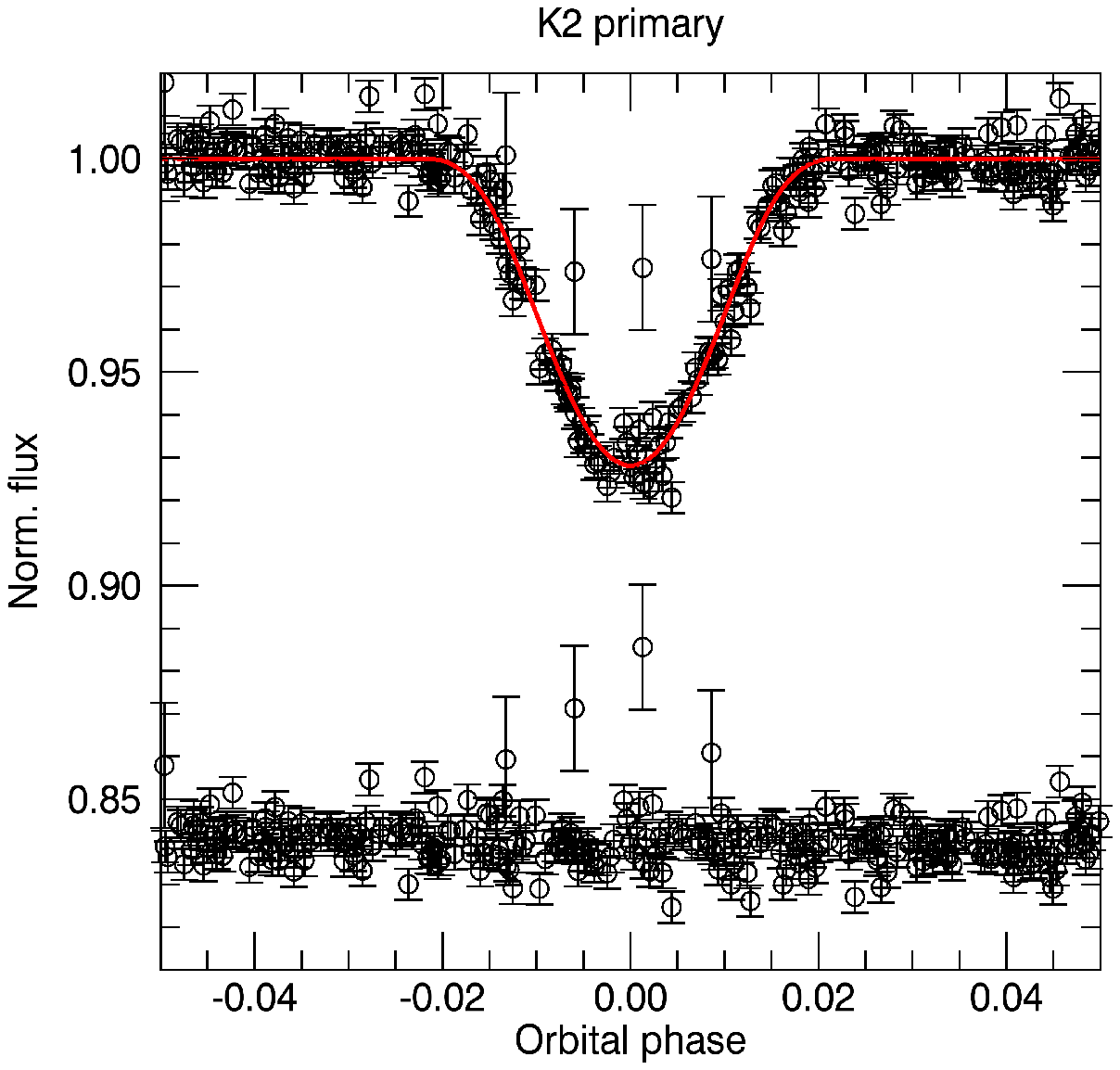}
  \includegraphics[width=0.49\linewidth, angle=0]{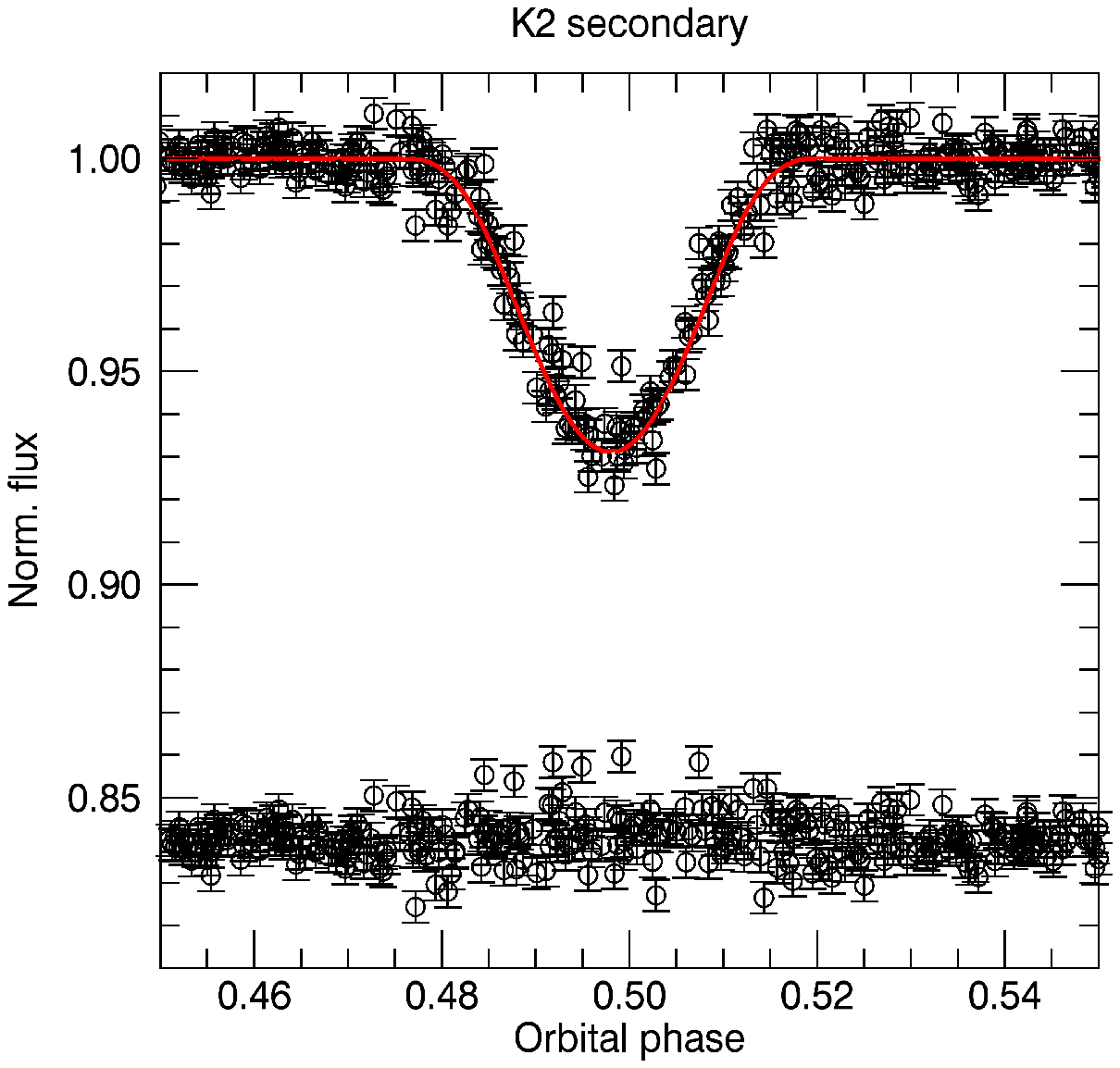}
   \caption{K2 light curves folded in phase and normalised to unity for the primary (left) 
and the secondary (right) component of USco16m25\@. 
The circles at the bottom of the plots represent the residuals between the fit (red line) 
and the observed transits (black circles with error bars).
The error bars were estimated from the out of eclipse dispersion of the K2 curve, as 0.37\%. 
The primary eclipse curve shows a few outliers during the eclipse phases; we increased the 
error bars of these data points (four) by an extra factor of 4, to reduce their impact in 
the final determined parameters.
The red line corresponds to the best-fit model to the global set
of data (light curves, photometric transits, and radial velocities).
}
   \label{fig_K2transitM5:K2_lc_in_phase}
\end{figure*}

The K2 light curve of USco16m25 shows eclipses with depths of about 8\% every 2.81 days, and out of eclipse 
photometric modulations with peak-to-peak amplitudes of up to 4\%. To prepare the eclipses for further studies 
described in this work, we attempted to minimise the effect of these modulations by normalising each individual 
eclipse using a parabolic fit to the regions immediate to the eclipse, as described in \citet{alonso08a}. 
Using our first estimate of the ephemeris of the eclipse, we removed the regions with eclipses and interpolated 
them to a regular grid of time points, to compute the autocorrelation of the out of eclipse data. This is 
plotted in Fig.\ \ref{fig_K2transitM5:K2lc_autocorrelation}, where the first maximum appears at a period close, 
but not equal to, the orbital period (2.6\,d vs. 2.8\,d). We attribute this modulation to stellar activity 
or clouds, in one or both components of the eclipsing binary. We also searched for effects at the orbital 
period (or its harmonics) of 
the eclipsing binary. These could be due to elliptical modulations due to tidal deformations of the stars, 
mutual reflection effects, Doppler beaming, or a combination of all. The light curve phased at the orbital 
period shows no sign of any of these effects with an amplitude larger than $\sim$0.2\%
(Fig.\ \ref{fig_K2transitM5:K2_lc_in_phase}).
We should also point out that we did not model the spots at the surface of the stars,
effect that can lead to unremoved systematics in the radial velocities and transit depth.

We derived the following ephemeris for the primary and the secondary from the K2 light curves
(the final adopted ones are discussed in Sect.\ \ref{K2transitM5:characterisation_parameters}):
\begin{itemize}
\item[$\bullet$] Primary (BJD) = 2456897.5228 $+$ 2.80882$\times$E
\item[$\bullet$] Secondary (BJD) = 2456896.1131 $+$ 2.80882$\times$E
\end{itemize}
where E stands for the number of eclipse.

Based on the ephemeris derived from the K2 light-curve (Section \ref{K2transitM5:K2_lc_phase}),
we undertook a large program of photometric and spectroscopic follow-up with a wide
battery of telescopes and instruments to confirm the nature of the transit 
and constrain the physical parameters of the system.

%
%
%

%
%
%

%
%
%

%
%
\begin{table}
 \centering
 \caption[]{Coordinates, photometry, and proper motion (mas/yr) of the K2 transiting system
quoted in Simbad as USco\,J161630.68$-$251220.1 (=\,EPIC\,203710387)\@. We list values from 
the USNO \citep{monet98,monet03}, Carlsberg Meridian Catalogue \citep{cmc14},
GAIA \citep{smart14}, 2MASS \citep{cutri03,skrutskie06}, UKIDSS \citep{lawrence07}, 
AllWISE \citep{wright10,cutri14}, PPMXL \citep{roeser10}, and NOMAD-1 \citep{zacharias05} 
catalogues. The original spectral type comes from  \citet*{luhman12c}, which we agree with in
this paper within the uncertainties.
}
 \begin{tabular}{l c} 
 \hline 
R.A.  &  16:16:30.68 \cr
dec   &  $-$25:12:20.1 \cr
$B_{J}$ (USNO A2.0)  & 19.5  \cr
$B_{J}$ (USNO B1.0)  & 19.70  \cr
$R1$ (USNO B1.0)  & 17.81  \cr
$R2$ (USNO B1.0)  & 17.54  \cr
SDSS$r$ (CMC14)  & 17.285  \cr
$I$ (USNO B1.0)  & 15.19  \cr
$G$ (GAIA) & 17.59$\pm$0.40 \cr
$I$ (DENIS)  & 15.042$\pm$0.07 \cr
$J$ (DENIS)  & 12.928$\pm$0.07 \cr
$K$ (DENIS)  & 11.916$\pm$0.09 \cr
$J$ (2MASS)   &  12.960$\pm$0.033 \cr
$H$ (2MASS)   &  12.282$\pm$0.030 \cr
$K_{s}$ (2MASS)   &  11.910$\pm$0.030 \cr
$Z$ (UKIDSS GCS)  &  14.208$\pm$0.003 \cr
$Y$ (UKIDSS GCS)  &  13.549$\pm$0.002 \cr
$J$ (UKIDSS GCS)  &  12.887$\pm$0.001 \cr
$H$ (UKIDSS GCS)  &  12.320$\pm$0.002 \cr
$K$ (UKIDSS GCS)  &  11.935$\pm$0.001 \cr
$w1$  & 11.748$\pm$0.023 \cr
$w2$  & 11.483$\pm$0.022 \cr
$\mu_{\alpha}\cos{\delta}$ (NOMAD) &  $-$8.0$\pm$4.0 \cr
$\mu_{\delta}$ (NOMAD) & $-$18.0$\pm$1.0 \cr
$\mu_{\alpha}\cos{\delta}$ (PPMXL) &  $-$11.8$\pm$5.1 \cr
$\mu_{\delta}$ (PPMXL) & $-$28.0$\pm$5.1 \cr
$\mu_{\alpha}\cos{\delta}$ (GCS) & $-$13.7$\pm$2.5 \cr
$\mu_{\delta}$ (GCS) & $-$20.0$\pm$2.5 \cr
SpType (literature) & M5 \cr
 \hline
 \label{tab_K2transitM5:properties_M5_primary}
 \end{tabular}
\end{table}
%

%
%
\section{Photometric follow-up with IAC80/CAMELOT}
\label{K2transitM5:phot_followup}
%

%
%
\begin{figure*}
  \centering
  \includegraphics[width=0.49\linewidth, angle=0]{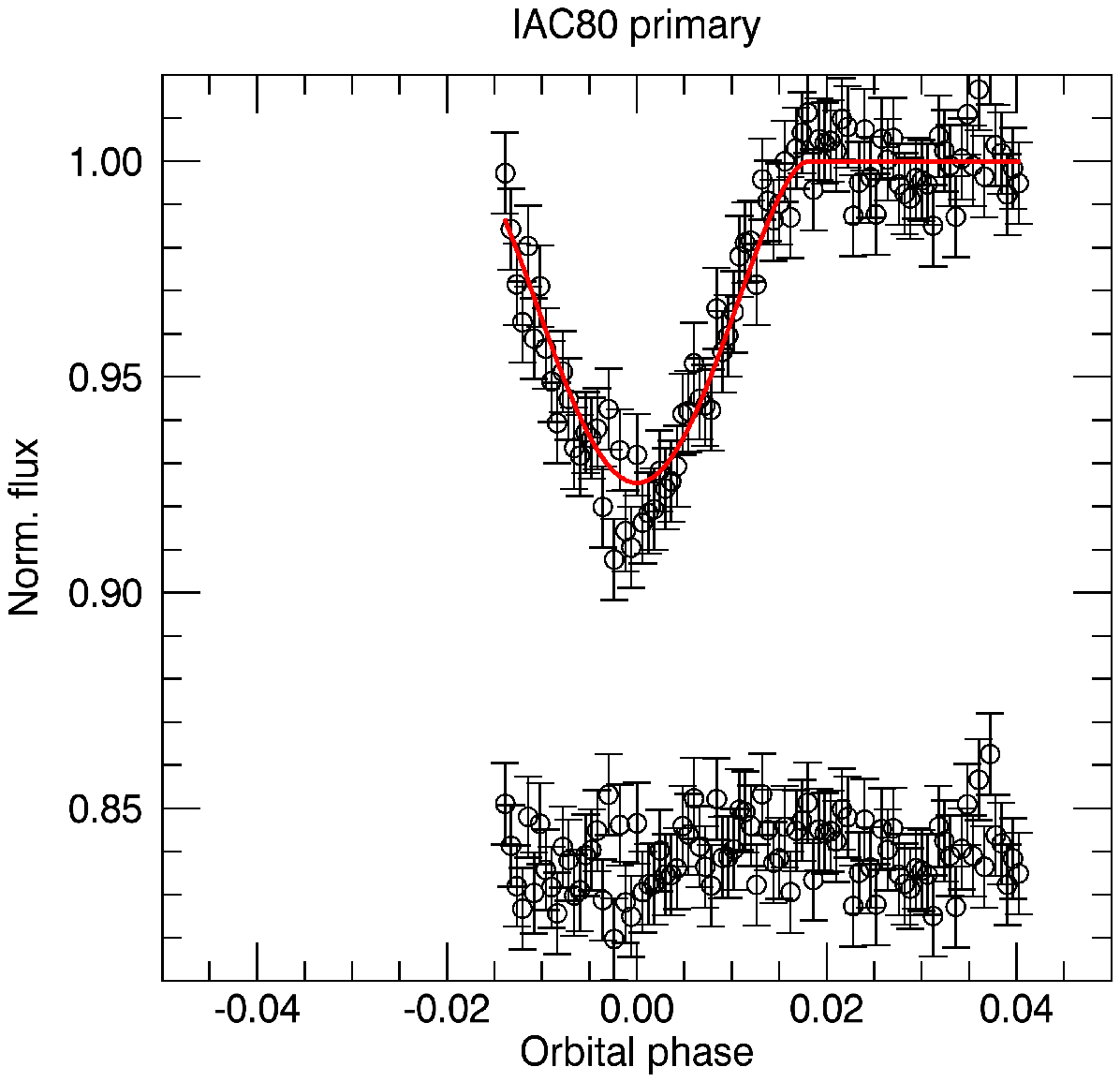}
  \includegraphics[width=0.49\linewidth, angle=0]{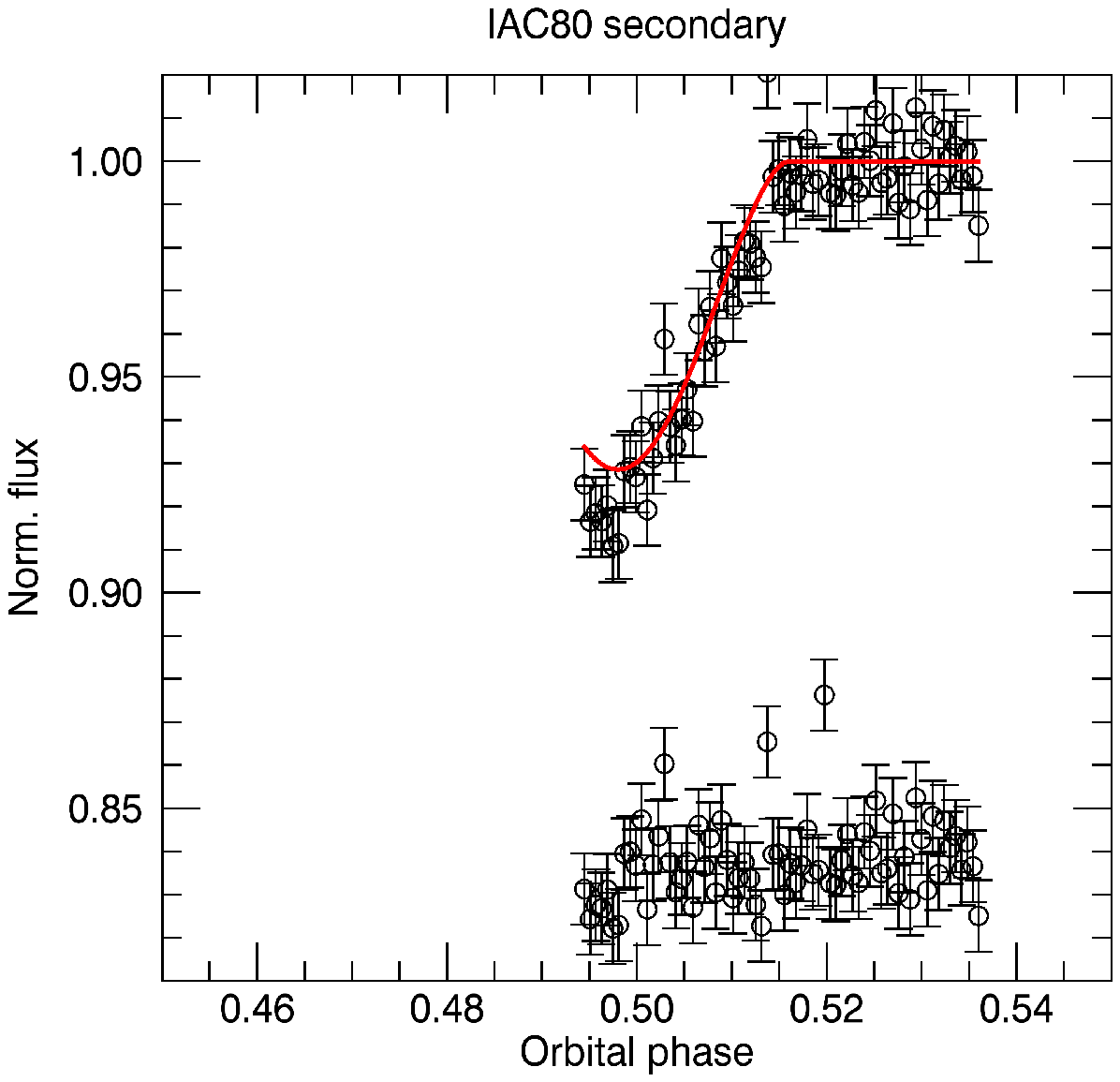}
   \caption{Two additional transits with different levels of completeness for USco16m25\@. 
We observed the system on the night of 18 May and 11 June 2015 in the SDSS$i$ filter 
with IAC80/CAMELOT\@. The red line corresponds to the best-fit model to the global set
of data (light curves, photometric transits, and radial velocities).
   }
   \label{fig_K2transitM5:additional_transits}
\end{figure*}

We attempted to observe several ground-based transits of USco16m25 with the CAMELOT instrument
(acronym for 'CAmara MEjorada LIgera del Observatorio del Teide' in Spanish) mounted on the 80-cm 
telescope IAC80 located at the Observatorio del Teide on the island of Tenerife in the 
Canary Islands (Spain). 

CAMELOT is a e2v 2048$\times$2048 back-illuminated Charged-Coupled device, providing a
field-of-view of 10.4$\times$10.4 arcmin with a pixel scale of 0.304 arcsec.
We used a read-out speed of 200 kHz, resulting in a read-out duration of 21.5 sec.

The presence of ``calima'' (sand from the Sahara) did not permit any observations on our first attempt 
on 11 May. However, we had two successful runs on the nights of 18 May and 11 June 2015, as described 
below. On the night of 18 May 2015, we obtained 91 frames with an on-source exposure 
time of 120\,s, from UT\,=\,00:24 until UT\,=\,04:03 UT\@. The average seeing, as measured from the 
frames, was 0.7 arcsec. These observations covered a partial primary eclipse event. On the night of 
11 June 2015, a secondary eclipse was partially observed, with 70 frames collected between UT\,=\,21h58 
and UT\,=\,00h46, with a mean seeing of 0.9 arcsec . After standard calibrations of the frames 
(removal of flat and bias contributions), we performed simple aperture photometry on USco16m25 and 
eight stars in the field of view. We used an aperture radius of 16 pixels. We selected the best four and
five stars for the first and second night, respectively, to produce an average reference star and obtain the 
final light curves plotted in Fig.\ \ref{fig_K2transitM5:additional_transits}.

%
%
\begin{figure*}
  \centering
  \includegraphics[width=0.49\linewidth, angle=0]{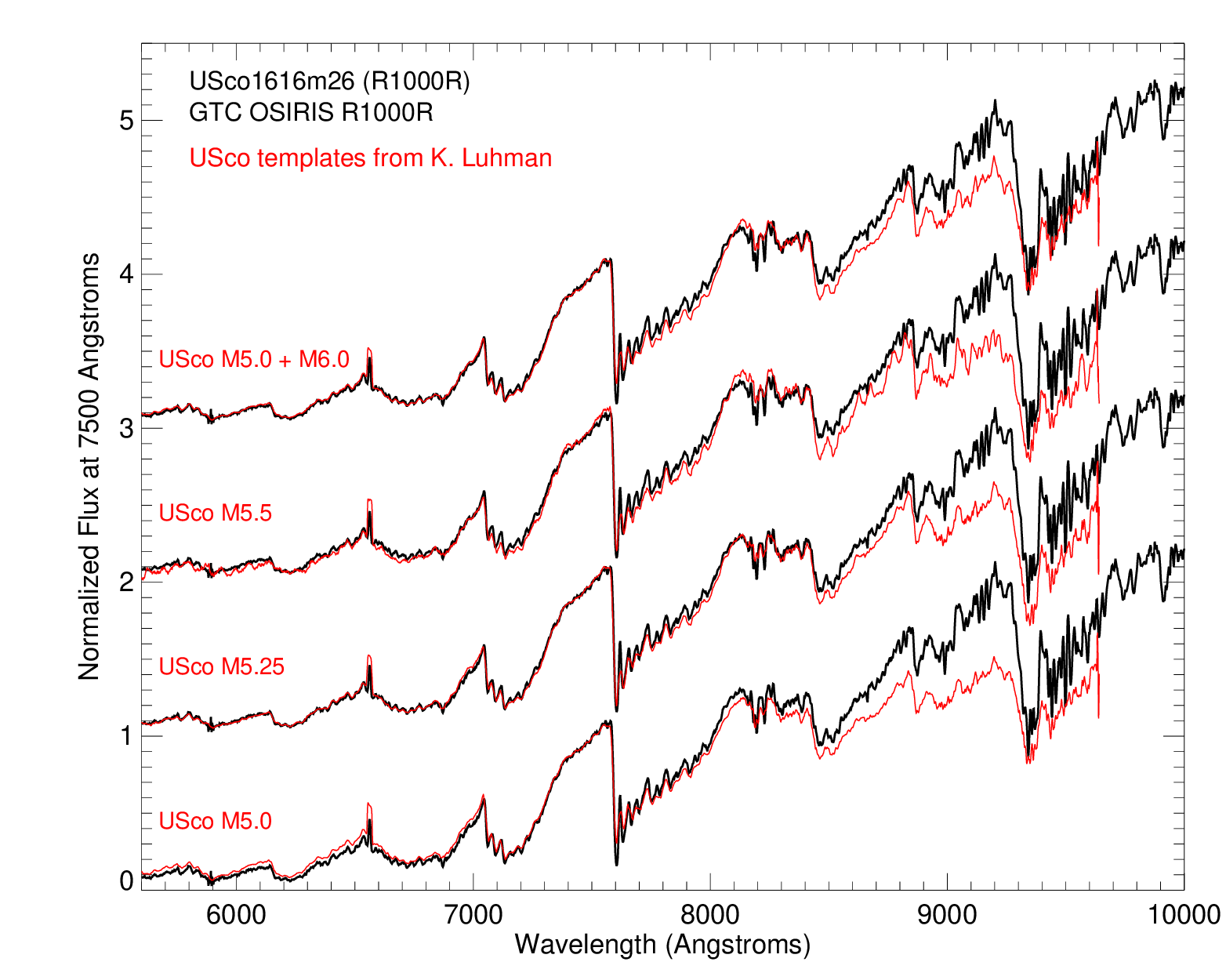}
  \includegraphics[width=0.49\linewidth, angle=0]{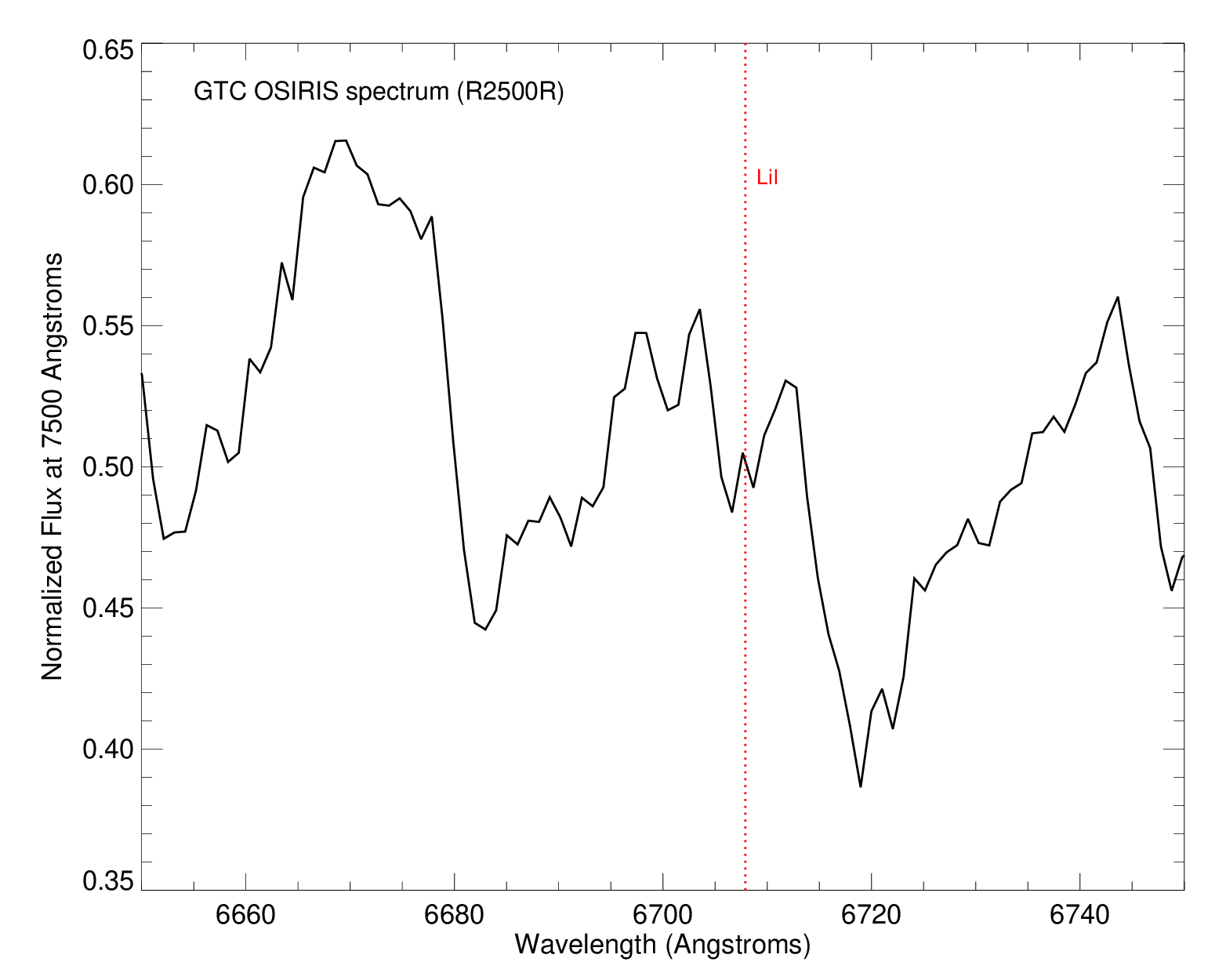}
   \caption{
{\it{Left:}} Optical spectra of USco16m25 (black) obtained with the R1000R grating on
GTC/OSIRIS\@. Overplotted in red are USco spectral template of spectral type M5, M5.25,
and M5.5 kindly provided by Kevin Luhman. The top spectrum represents a combination
of the normalised spectra of a M5 and a M6 member of USco.
{\it{Right:}} Zoom-in on the lithium absorption line at 6707.8\AA{} (marked with a vertical
dashed red line) of USco16m25 (black) obtained with the R2500R grating on GTC/OSIRIS\@.
}
   \label{fig_K2transitM5:opt_spec_GTC}
\end{figure*}
%

%
%
\section{Spectroscopic follow-up}
\label{K2transitM5:spec_followup}
\subsection{Long-slit spectroscopy}
\label{K2transitM5:spec_followup_longslit}
\subsubsection{GTC/OSIRIS optical spectra}
\label{K2transitM5:spec_GTC_OSIRIS}

We obtained low-resolution optical spectroscopy with OSIRIS
\citep[Optical System for Imaging and low-intermediate Resolution Integrated Spectroscopy;][]{cepa00}
mounted on the 10.4-m GTC telescope in La Palma. We observed the target on 
20 May 2015 (UT\,=\,01h56m) in service mode as part of a filler program GTC38-15A (PI Lodieu).
The system was in phase 0.725, so we decided to re-observe it with the grating R1000R 
in phase 0 when the light from the secondary is mostly blocked (2 August 2015 at UT\,=\,21h20m) 
and at phase 0.5 when the light of the primary is blocked (9 August 2015 at UT\,=\,21h45m) 
to compare the combined spectra and the contribution from each component because eclipses 
are not total (i.e.\ grazing). Unfortunately, bad weather in La Palma on 9 August 2015 did not allow 
us to fully characterise the degree of light from the secondary in the optical spectrum. No
other attempt was made after that date because USco was too low for observations in this 
configuration from La Palma.
 
OSIRIS is equipped with two 2048$\times$4096 Marconi CCD42-82 detectors offering 
a field-of-view approximately 7$\times$7 arcmin with a binned pixel scale of 0.25 arcsec.
USco16m25 was observed between UT\,=\,01h55 and UT\,=\,02h35 when USco was the
highest in the Canarian sky, under dark conditions and a seeing of $\sim$1.6 arcsec.
We used a slit of 1.0 arcsec. We obtained a single exposure of 600\,sec with the R1000R
grating (510--1000 nm) and three exposures of 300\,sec shifted along the slit with the 
R2500R grating (557.5--768.5 nm).
Bias frames, dome flat fields, as well as Neon, Xenon, and HgAr arc lamps were observed 
by the observatory staff during the afternoon.

We reduced the OSIRIS optical spectrum under the IRAF environment 
\citep{tody86,tody93}\footnote{IRAF is distributed by the National Optical Astronomy 
Observatory, which is operated by the Association of Universities for Research in Astronomy
(AURA) under a cooperative agreement with the National Science Foundation.}.
We subtracted the raw spectrum by a median-combined bias and
divided by a normalised median-combined dome flat field taken during the
afternoon. We extracted optimally a 1D spectrum from the 2D image and
calibrated that spectrum in wavelength with the lines from the combined arc lamp.
We corrected our spectra for instrumental response with the Ross\,640 spectro-photometric standard 
\citep[DZ5;][]{monet03,lepine05d,sion09}. In the case of the R1000R grating
we created a response function corrected for the second-order contamination 
beyond $\sim$9200\AA{} by joining the response of the detector
from the observations of the standard star with the R1000R grating and
the same grating in combination with the Sloan $z$ broad-band filter.
The final GTC/OSIRIS spectrum of USco\,J16163068$-$2512201, normalised at 750nm, 
is displayed in the left-hand side panel in Fig.\ \ref{fig_K2transitM5:opt_spec_GTC}
along with known USco members.

\subsubsection{Subaru/FOCAS optical spectra}
\label{K2transitM5:spec_Subaru_FOCAS}

We obtained low-resolution optical spectra with the FOCAS 
\citep[Faint Object Camera and Spectrograph;][]{kashikawa02} instrument mounted on the 8.2m 
Subaru telescope in Hawaii. We observed USco16m25 on 19 June 2015 (corresponding to a phase of 
$\sim$0.13) during unoccupied times of Subaru telescope normal program (PI: Y.\ Kawashima).

FOCAS is equipped with two 2048$\times$4096 CCD detectors offering a circular field-of-view 
of 6 arcmin in diameter with the pixel scale of 0.1 arcsec. We observed USco16m25 between 
UT\,$=$\,07h20m and UT\,$=$\,08h24m when Upper Scorpius was high in the Hawaiian sky, under 
dark conditions but persistent cirrus with a seeing of $\sim$1.5 arcsec. We used a long slit 
with a slit width of 0.4 arcsec. We obtained 10 exposures of 600\,sec with the 300B grating 
and the SY47 filter, providing the spectral resolution of R$\sim$1000 over the 470--910 nm
wavelength coverage.

\begin{figure}
  \centering
  \includegraphics[width=\linewidth, angle=0]{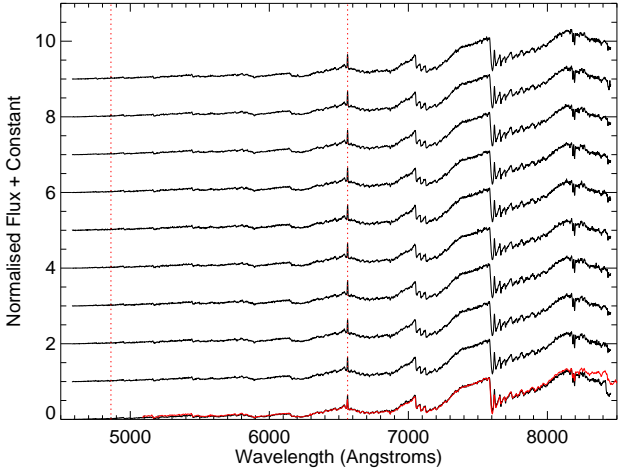}
   \caption{
Optical spectra of USco16m25 (black) obtained with Subaru/FOCAS at a resolution of 1000\@.
The spectra were obtained on 19 June 2015 between UT\,=\,07h and UT\,=\,08h24\@.
Overplotted in red on the bottom spectrum is the GTC/OSIRIS spectrum obtained on
2 August 2015 at a similar resolution (see left-hand side panel of Fig.\ \ref{fig_K2transitM5:opt_spec_GTC}.
The red dotted vertical lines mark the positions of H$\beta$ and H$\alpha$ lines
at 4861\AA{} and 6563\AA{}, respectively.
}
   \label{fig_K2transitM5:opt_spec_Subaru}
\end{figure}

The spectra are de-biased, flat-fielded, traced, sky-subtracted, and extracted with standard 
IRAF procedures. Wavelength calibration is done using Th-Ar lines. Sensitivity calibration is 
not applied, because of a lack of photometric standard observation on that night and large 
differences of sky conditions for photometric standard data taken on other nights. We note that 
the absence of sensitivity calibration has less impact on overall results of this paper, and 
the spectra taken with Subaru/FOCAS well match with the ones obtained with GTC/OSIRIS
(Fig.\ \ref{fig_K2transitM5:opt_spec_Subaru}). The similarity between the FOCAS and
OSIRIS spectra also demonstrates that the contribution of the light from the secondary
is extremely small up to 850 nm or that both objects have very similar spectra energy
distribution at these wavelengths, statement consistent with the similarity observed
between the GTC spectra taken at phase 0 and phase 0.725\@.

%
%
%

%
%
\begin{table*}
 \centering
 \caption[]{Logs of WHT/ISIS spectroscopic observations for the target, the 
 telluric standard star HIP77859, and the RV templates HD\,164922
 and GJ\,699 (=Barnard's star). The UT corresponds to the middle times of the
 respective on-source integrations.
}
 \begin{tabular}{c c c c c c c} 
 \hline 
Target  & Date  &  UT  & Grating & Slit  & ExpT  & Airmass \cr
 \hline
            & DDMMYYYY & hh:mm:ss.ss &        &  arcsec & sec &      \cr 
 \hline
USco16m25   & 02062015  & 23:29:05.6 & R1200R & 1.2  & 1800  & 1.83  \cr 
            & 03062015  & 00:09:12.1 & R1200R & 1.2  & 1800  & 1.73  \cr 
            &           & 00:42:51.8 & R1200R & 1.2  & 1800  & 1.70  \cr 
            &           & 01:54:40.2 & R1200R & 1.2  & 1800  & 1.81  \cr 
HD\,164922  & 02062015  & 23:14:04.9 & R1200R & 1.2  &   10  & 1.XX  \cr 
 \hline
USco16m25   & 09062015  & 00:22:51.9 & R1200R & 1.2  & 1800  & 1.70  \cr 
            &           & 00:56:44.9 & R1200R & 1.2  & 1800  & 1.73  \cr 
            &           & 01:57:26.6 & R1200R & 1.2  & 1800  & 1.93  \cr
            &           & 02:30:04.9 & R1200R & 1.2  & 1800  & 2.15  \cr 
            &           & 03:02:22.3 & R1200R & 1.2  & 1800  & 2.51  \cr 
HIP\,77859  & 09062015  & 01:32:30.0 & R1200R & 1.2  &   30  & 1.86  \cr 
HIP\,77859  & 09062015  & 01:36:30.0 & R1200R & 1.2  &   30  & 1.88  \cr 
GJ\,699     & 09062015  & 03:22:00.0 & R1200R & 1.2  &   60  & 1.17  \cr 
GJ\,699     & 09062015  & 03:24:00.0 & R1200R & 1.2  &   60  & 1.17  \cr 
 \hline
USco16m25   & 17062015  & 21:45:07.6 & R1200R & 1.2  & 1800  & 2.08  \cr
HIP\,77859  & 17062015  & 21:14:59.0 & R1200R & 1.2  &   10  & 2.06  \cr 
GJ\,699     & 17062015  & 22:11:53.4 & R1200R & 1.2  &   40  & 1.62  \cr 
 \hline
USco16m25   & 07072015  & 22:56:45.0 & R1200R & 1.2  & 1800  & 1.72  \cr
HIP\,77859  & 07072015  & 23:22:13.5 & R1200R & 1.2  &   30  & 1.79  \cr
GJ\,699     & 07072015  & 23:39:44.0 & R1200R & 1.2  &   60  & 1.10  \cr
 \hline
USco16m25   & 12072015  & 22:03:48.9 & R1200R & 1.2  & 1800  & 1.70  \cr
HIP\,77859  & 12072015  & 22:26:13.6 & R1200R & 1.2  &   10  & 1.69  \cr
GJ\,699     & 12072015  & 22:38:36.9 & R1200R & 1.2  &   30  & 1.14  \cr
 \hline
USco16m25   & 13072015  & 22:01:44.0 & R1200R & 1.2  & 1800  & 1.70  \cr
HIP\,77859  & 13072015  & 22:22:25.5 & R1200R & 1.2  &   10  & 1.69  \cr
GJ\,699     & 13072015  & 23:39:44.0 & R1200R & 1.2  &   60  & 1.10  \cr
 \hline
USco16m25   & 15072015  & 00:05:36.1 & R1200R & 1.2  & 1800  & 2.13  \cr
HIP\,77859  & 15072015  & 00:27:06.1 & R1200R & 1.2  &   10  & 2.53  \cr
GJ\,699     & 15072015  & 00:35:10.8 & R1200R & 1.2  &   20  & 1.13  \cr
 \hline
 \label{tab_K2transitM5:log_ISIS_spectra}
 \end{tabular}
\end{table*}
\subsection{Radial velocity measurements with WHT/ISIS}
\label{K2transitM5:RV_WHT_ISIS}

We obtained several sets of optical spectra with the ISIS (Intermediate dispersion Spectrograph
and Imaging System) instrument on the Cassegrain focus of the 4.2m William Herschel Telescope
(WHT) over the course of several nights (Table \ref{tab_K2transitM5:log_ISIS_spectra}).

ISIS is a double-armed medium-resolution spectrograph equipped with a dichroic to spread the
light in a blue and red arm. We are mainly interested in the red arm for our purposes because the
target is a mid-M with little flux short wards of 500 nm. We used the same grating R1200R and the
1.2 arcsec slit covering the $\sim$745-845 nm wavelength range at a spectral resolution of
approximately 6000\@. All spectra were taken at parallactic angle.
Arc lamps of CuAr$+$CuNe were observed immediately after the target
while bias and flat field frames were obtained during the afternoon preceding both nights.
We used different grating in the blue arm (R600B, R1200B, and R300B) during the first three
sets of observations, confirming the low level of flux and lack of strong emission.

On the night of 2 June 2015 we took four spectra of USco16m25 as part of service programme number
SW2015a31 (PI Lodieu). We also obtained six spectra of the radial velocity (RV) standard star HD164922
\citep[G9V; $I$\,=\,6.1 mag; RV\,=\,20.29$\pm$0.09 km/s;][]{nidever02,gray03,vanLeeuwen07}.
This night was clear with variable seeing but the moon was located at 13--14 degrees from the
target, resulting in one the worst quality spectra among all obtained with ISIS\@. We obtained
another five spectra of USco16m25 on the night of 8 June 2015 as part of a DDT
program to continue our RV follow-up of the system. The moon rose up at UT\,=\,1h
with an illumination of 56\% and was 94 degrees away from USco. We also observed the telluric
standard HIP\,77859 \citep[B2V; $V$\,=\,5.4 mag; RV\,=\,$-$9.20$\pm$3.1 km/s;][]{houk88,gontcharov06,vanLeeuwen07}
and Barnard's star as a RV standard because of its spectral type comparable to our
target \citep[M4; $I$\,=\,6.74; RV\,=\,$-$110.5$\pm$0.1 km/s;][]{nidever02,gizis97a,vanLeeuwen07,koen10a}.
This night was affected by thin cirrus but the seeing was sub-arcsec at zenith.
From this night on, we also observed the telluric standard HIP\,77859 and the Barnard's star
as a RV standard using the same configuration. We obtained another spectrum of
the system on 17 June thanks to the collaboration with the team leading the ITP programme
ITP13-8 (PI Rubi\~no-Martin). This night was clear with seeing below 0.8 arcsec although most
of the observations were obtained during twilight. We collected three additional points for
our target with the same configuration and same standard stars on 12, 13, and 14 July 2015
under variable conditions mainly affected by dust from the Sahara.
The log of all WHT/ISIS observations is provided in Table \ref{tab_K2transitM5:log_ISIS_spectra}.

We reduced the ISIS optical spectra under IRAF\@.
We subtracted each raw spectrum by a median-combined bias and divided by a normalised
median-combined dome flat field taken during the afternoon. We extracted optimally the
spectrum and calibrated it in wavelength with the lines from the CuAr$+$CuNe arc lamps.
We did not apply any flux calibration since we are only interested in the RV\@.
We corrected the spectra for telluric lines with the telluric standard star HIP\,77859
for all nights, except for the data taken on 2 June where no telluric standard was observed.
Hence, we corrected for telluric lines with the standard observed on 7 July at a similar
airmass but we can see that the correction is the poorest of all.
Nonetheless, we can resolve each line of the sodium doublet, showing that our target
is indeed a spectroscopic binary (Fig.\ \ref{fig_K2transitM5:opt_spec_ISIS}).
The sequence of ISIS spectra of USco16m25 is shown in Fig.\ \ref{fig_K2transitM5:opt_spec_ISIS}.

%
%
\begin{figure*}
  \centering
  \includegraphics[width=0.49\linewidth, angle=0]{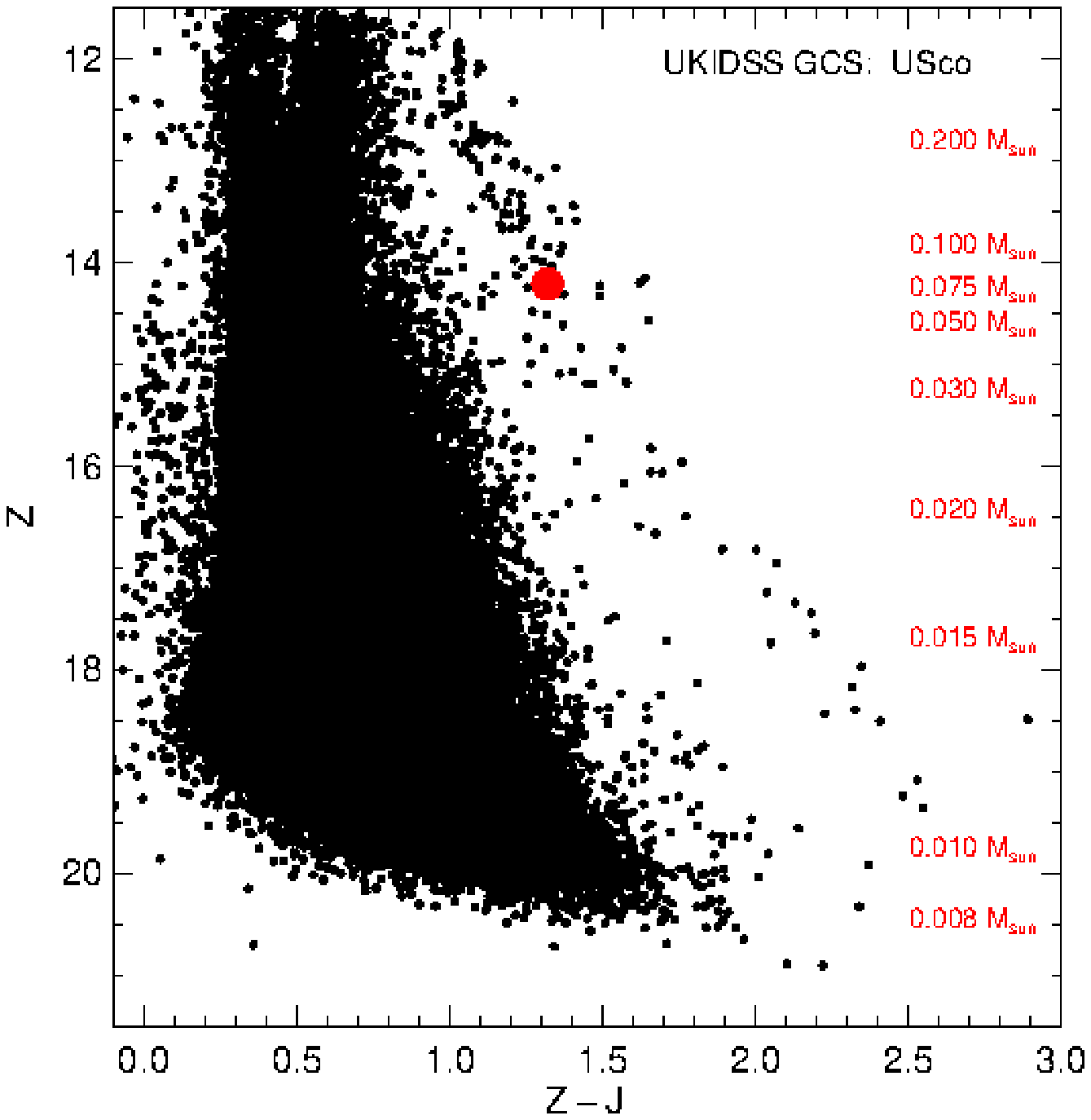}
  \includegraphics[width=0.49\linewidth, angle=0]{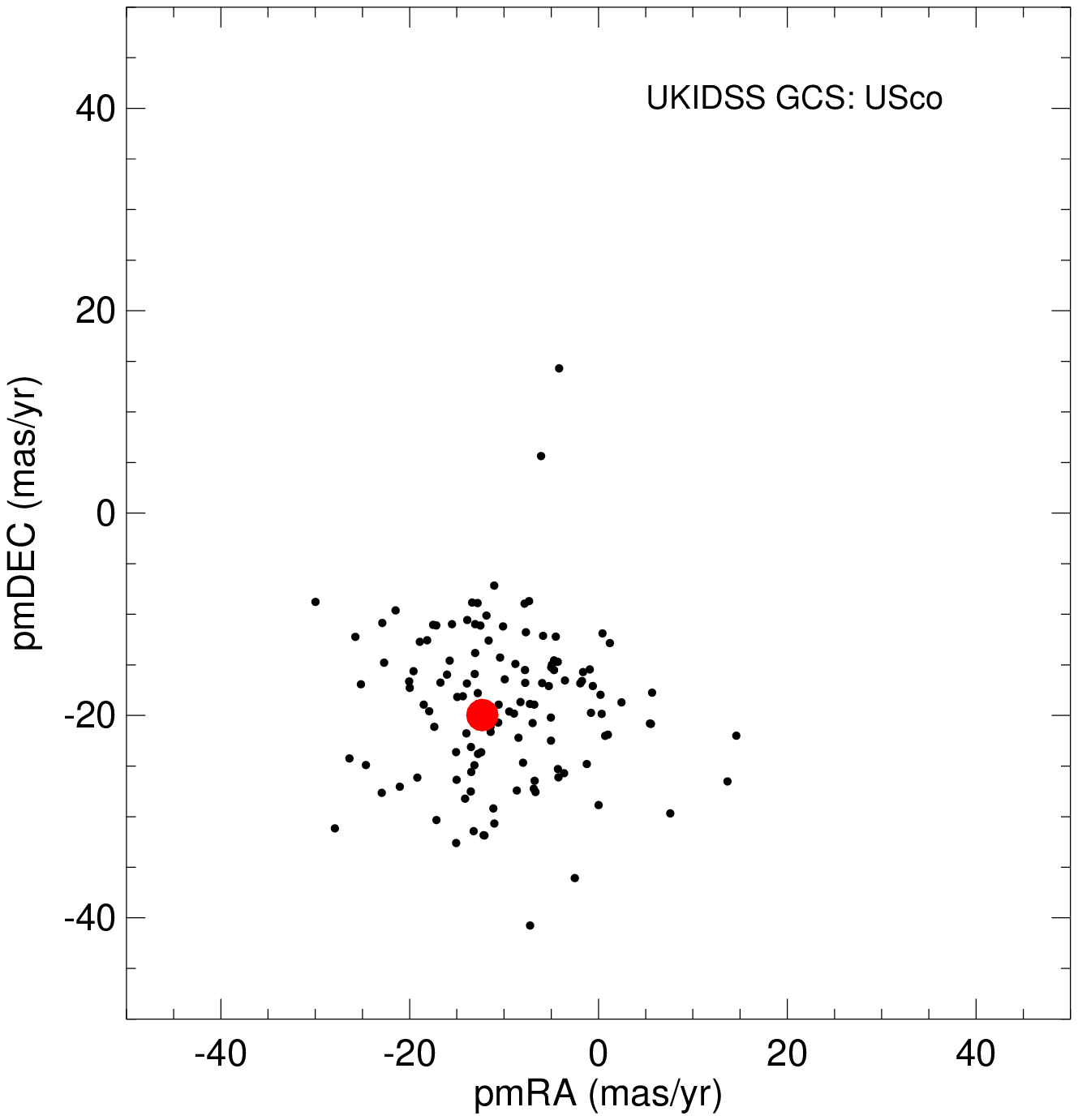}
  \caption{
{\it{Left:}} (Z-J,Z) colour-magnitude diagram for 6.5 deg$^{2}$ in the central region
of USco (Lodieu et al.\ 2007). The USco cluster sequence is well separated from field stars.
{\it{Right:}} Vector point diagram. We averaged the values of the proper motion in each
direction from the datasets listed in Table \ref{tab_K2transitM5:properties_M5_primary},
where the error bars represent the dispersion between each catalogue. The mean proper motion
of the association is ($\mu{_\alpha}\cos\delta$,$\mu_{\delta}$)\,$\sim$\,($-$11,$-$25) mas/yr.
Our transiting system is highlighted with a large red dot, demonstrating its photometric and
astrometric membership.}
\label{fig_K2transitM5:CMD_VPD}
\end{figure*}
%

%
%
\begin{figure*}
  \centering
  \includegraphics[width=0.49\linewidth, angle=0]{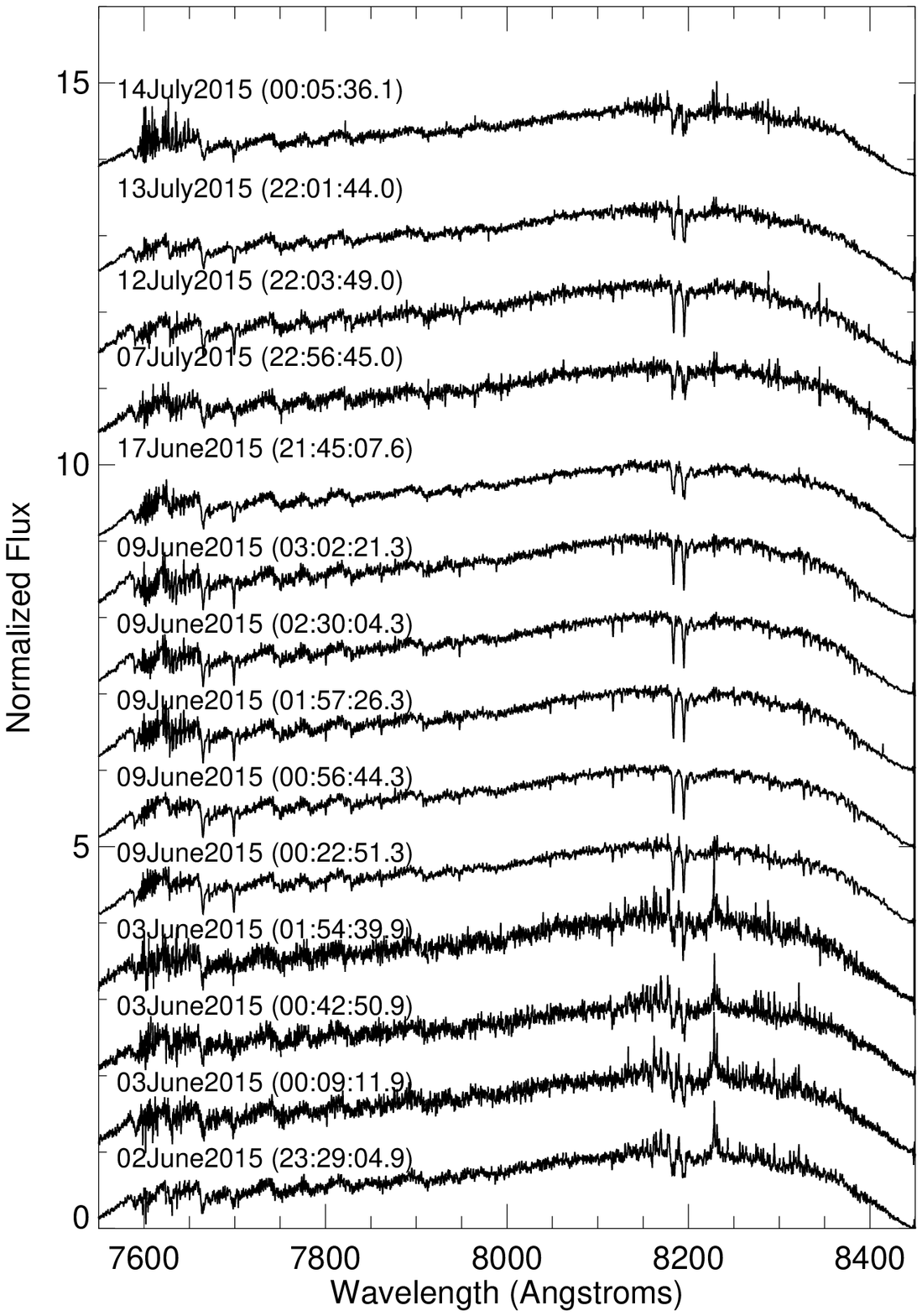}
  \includegraphics[width=0.49\linewidth, angle=0]{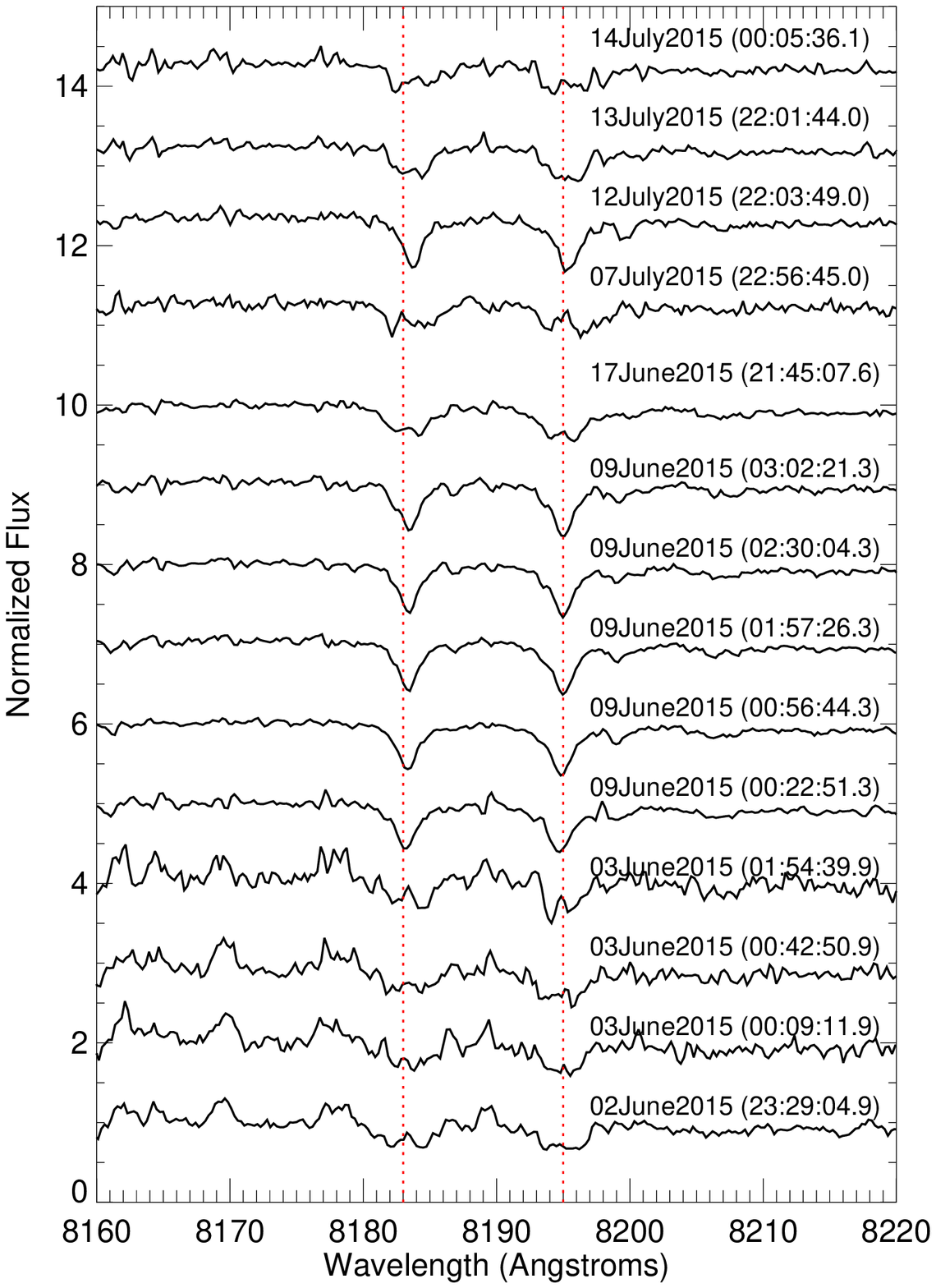}
   \caption{Sequence of optical spectra of USco16m25 taken with WHT/ISIS over the
   course of 1.5 month, between 2 June and 14 July 2015\@. The wavelength range spans
   745--845 nm at a spectral
   resolution of $\sim$6000 (left plot) which includes the gravity-sensitive sodium
   (818.3/819.5 nm; right-hand side plot) doublets at 818.3/819.5 nm.
}
   \label{fig_K2transitM5:opt_spec_ISIS}
\end{figure*}
%

%
%
\section{Characterisation of the transiting system}
\label{K2transitM5:characterisation}
\subsection{Membership to USco}
\label{K2transitM5:characterisation_USco_memb}

The system is a photometric and astrometric member of the USco association
(Fig.\ \ref{fig_K2transitM5:CMD_VPD}). It position in the ($Z-J$,$Z$) colour-magnitude diagram 
demonstrates that it lies on the USco sequence determined from the dataset of the UKIRT Infrared 
Deep Sky Survey \citep[UKIDSS;][]{lawrence07} Galactic Clusters Survey \citep{lodieu06,lodieu07a,lodieu13c}.
Its $J$-band magnitude is consistent with the spread of measurements for M5--M5.5
USco members \citep[e.g.][]{slesnick08,lodieu11a}.
USco16m25 also matches the position of known USco members in the vector point
diagram dispplayed on the right-hand side of Fig.\ \ref{fig_K2transitM5:CMD_VPD}
(see magnitude and proper motions reported in Table \ref{tab_K2transitM5:properties_M5_primary}).

We obtained a low-resolution spectrum of the system at phase 0 with GTC/OSIRIS 
(Sect.\ \ref{K2transitM5:spec_GTC_OSIRIS}), classified as a M5 by \citet*{luhman12c}.
We compared our spectrum to known USco members\footnote{We thank Kevin Luhman for 
kindly providing his templates prior to publication} and classified USco16m25 as M5.25$\pm$0.25
from the spectral region in the 550--850 nm interval, 
in close agreement with the determination by \citet*{luhman12c}. This suggests that the object 
lies at the stellar/substellar boundary which is set to M6.5 based on models \citep{luhman99a}. 
According to the scale of effective temperatures for young pre-main sequence dwarfs
\citep{luhman99a}, USco16m25 would have a temperature of approximately 3100$\pm$50\,K
\citep[see also][]{luhman03b,rice10,lodieu11a}. We also 
compared our spectrum with field dwarfs of similar spectral types from the Sloan database 
\citep{bochanski07a} and found worst matches than the one for members of $\eta$\,Cha 
over the 6000--9000\,\AA{} wavelength region.

We also measured the H$\alpha$ equivalent widths in each individual spectra, yielding
values between 6.2 and 6.7\,\AA{} over the course of 30\,min. Assuming a mean value 
of 6.5/2\,=\,3.25\AA{} for the equivalent width of H$\alpha$ of each component and applying 
equations of \citet{stassun12a}, chromospheric activity would imply a correction of 
10.7$\pm$5.8\% in the radius (inflation) and $-$5.0$\pm$2.3\% in the effective temperature 
(suppression). A correction for an inflated radius for direct comparison with non-active models 
would imply a older age for the system and the USco association, but we note that we did not 
apply this correction to our final results described in Sect.\ \ref{K2transitM5:characterisation_parameters}.

We also investigated the
strength of the gravity-sensitive sodium doublet at 8183/8195\,\AA{} in the lowest
resolution spectrum, implying pseudo-equivalent widths of 1.7/2.1\,\AA{} for each 
component of the doublet. These values agree with the mean values of M5 dwarf members 
in USco \citep[Fig.\ 4 of][]{lodieu11a}. The presence of H$\alpha$ in emission and
the weak sodium doublet confirm that the system is young and that the K2 light curve 
is really associated to the binary system.

%
%
\begin{table*}
\begin{center}
\caption{Radial velocities of USco16m25 derived from the cross-correlation function method.
RVs are estimated by performing a 2nd order polynomial fit to the peak of the CCFs.
F21 represents the secondary-to-primary flux ratio, F2/F1, given by the ratio
of the peaks of the cross-correlation functions of the secondary (F2) and primary (F1) component.
The error of this measurement is giving by eF21\@.
$^a$ Heliocentric julian day.
$^b$ Radial velocity of the primary.
$^c$ Radial velocity of the secondary.
$^d$ Discarded RV point during the RV fitting procedure.
}
\begin{tabular}{l c c c c c c c c}
\hline
\hline
Date & HJD$^{a}$ &  Phase & RV$_{1}^{b}$  & $\delta$RV$_{1}$ &   RV$_{2}^{c}$  & $\delta$RV$_{2}$ & F21 & eF21 \\
ddmmyyyy        & days        &   & km s$^{-1}$  &  km s$^{-1}$ & km s$^{-1}$  &  km s$^{-1}$  &  & \\
\hline
03062015 & 2457176.484357799 & 0.300   & $-$36.29       &  0.33   & $+$38.83   &  0.32 & 0.9664 & 0.0481 \\
03062015 & 2457176.512210492 & 0.314   & $-$31.37       &  0.43   & $+$36.85   &  0.43 & 0.9818 & 0.0765 \\
03062015 & 2457176.535586138 & 0.332   & $-$40.79       &  0.47   & $+$32.96   &  0.68 & 0.9906 & 0.1018 \\
03062015 & 2457176.585451550 & 0.350   & $-$25.71       &  0.47   & $+$36.41   &  0.47 & 0.9813 & 0.0687 \\
09062015 & 2457182.521620525 & 0.463   & $-$0.56        &  0.12   & ---        & ---   &  ---   &  ---   \\
09062015 & 2457182.545150519 & 0.472   & $-$1.17        &  0.10   & ---        & ---   &  ---   &  ---   \\
09062015 & 2457182.587298515 & 0.487   & $+$0.33        &  0.12   & ---        & ---   &  ---   &  ---   \\
09062015 & 2457182.609963939 & 0.495   & $+$1.67        &  0.13   & ---        & ---   &  ---   &  ---   \\
09062015 & 2457182.632386450 & 0.503   & $+$4.42        &  0.14   & ---        & --    &  ---   &  ---   \\
17062015 & 2457191.411856342 & 0.629   & $+$27.95       &  0.14   & $-$28.81   &  0.21 & 0.9631 & 0.0357 \\
07072015 & 2457211.460653753 & 0.767   & $+$45.23       &  0.29   & $-$42.54   &  0.31 & 0.9634 & 0.0247 \\
12072015 & 2457216.423575887 & 0.534   & $-$1.92$^d$    &  0.17   & ---        & --    &  ---   &  ---   \\
13072015 & 2457217.422062478 & 0.889   & $+$25.33       &  0.19   & $-$25.73   &  0.24 & 0.9638 & 0.0525 \\
14072015 & 2457218.508007215 & 0.276   & $-$39.45       &  0.25   & $+$43.86   &  0.29 & 0.9579 & 0.0562 \\
\hline
\end{tabular}
\label{tab_K2transitM5:ccfrv}
\end{center}
\end{table*}

In the highest resolution spectrum, we resolved the lithium absorption line at 6707.8\,\AA{} 
into two components because the target was observed at a phase of 0.725\@. We measured 
pseudo-equivalent widths for the lithium line of 0.19$\pm$0.1 and 0.29$\pm$0.1\,\AA{} for the primary 
and the secondary, respectively (right-hand side panel in Fig.\ \ref{fig_K2transitM5:opt_spec_GTC}).
The individual values, which should be corrected for the dilution factor, agree with the
pseudo-equivalent widths reported by \citet{kraus15a} for the M4.5 eclipsing binary and earlier
studies of the USco population \citep{preibisch01,preibisch02a,lodieu11a,rizzuto15},
adding further credit to the membership of USco16m25\@. Our measurements of the lithium 
pseudo-equivalent width of the secondary translate into an abundance of $\log$N(Li)\,=\,3.1$\pm$0.3 
(in the usual scale of $\log$N(H)\,=\,12), consistent with no depletion
\citep[see Figure 16 of][]{zapatero02c}. For the primary, we would infer an abundance of
$\log$N(Li)\,=\,2.0$\pm$0.4, which suggests some lithium destruction may have taken place in 
this slightly more massive object. While the absolute error bars on the equivalent widths consistent
with other M dwarfs in USco prevent us to reach any strong 
conclusion, it is worth mentioning that the differential error in the measurement is smaller and
this result could be taken as an indication that in the primary star, presumably a fully convective 
object, the core temperature has already reached the temperature needed to ignite lithium burning 
via $^7$ Li\,$+$\,p\,$\rightarrow$\,$\alpha$ (T$\sim$2.5$\times$10$^6$\,K) while in the secondary 
this is not the case yet. We note that interior evolution models \citep[e.g.][]{baraffe15} predict
lithium burning temperatures are reached at significantly older ages ($>$30 Myr) at 
0.09 M$_{\odot}$. More accurate lithium observations are worth to be pursued as they will
bring insight on the properties of the matter in the interior of these degenerate objects with 
potential implications on the equation of state or opacities adopted.

\subsection{Radial velocity confirmation}
\label{K2transitM5:characterisation_RV}

Figure \ref{fig_K2transitM5:opt_spec_ISIS} shows the full WHT/ISIS spectra on the left-hand
side with a zoom on the potassium (K{\small{I}} and sodium (Na{\small{I}}) doublets.
We observe that both doublets are resolved on several nights with a change in the
position of individual lines, e.g.\ on the night of 13 July (phase 0.889) and 14 July 
(phase 0.276). We measured the offsets of the Na{\small{I}} doublet with respect to 
the air wavelength\footnote{Central wavelengths of the lines of the doublet (8183.25\AA{} 
and and 8194.79\AA{} are taken from the NIST (National Institute of Standards and Technology) 
atomic spectra database lines} with task {\tt{splot}} in IRAF\@. On the night of 9 June
and 12 July 2015, the doublet was unresolved because the observations took place very
close to transit (phase between 0.46 and 0.53). 

The RV measurements were extracted from the ISIS spectra (R$\sim$6000) 
using the cross-correlation function (CCF) technique, within the package 
MOLLY\footnote{The MOLLY package can be download from 
www2.warwick.ac.uk/fac/sci/physics/research/astro/people/marsh/software/}.
We first normalised the ISIS spectra of the target and the template star 
GJ699 (Barnard's star) within the IRAF package using a low-order legendre
polynomial. Then, we cross-correlated each target spectrum with the template
stellar spectrum. We did not get a spectrum of the template star every night
(see Sect.\ \ref{K2transitM5:RV_WHT_ISIS}), thus, we used the spectrum of 
the night taken on 09 June 2015\@. We also corrected for possible drifts using the
spectra of the sky subtracted during the reduction procedure. We derived the
drifts by cross-correlating the sky spectrum of every target spectra with the sky 
spectrum of the template star taken on 17 July 2015 \@. We found drifts in the 
range $-$0.8 km/s to 3.7 km/s.

In Table\ \ref{tab_K2transitM5:ccfrv}, we provide the heliocentric julian days
and radial velocities for both components of the system. We do not see the double
peak CCF in the spectra taken at orbital phases close to 0.5\@. For those cases, 
we associated these RVs to the primary star. The RV uncertainties are in the range 
0.1--0.7 km/s. We fitted these RVs with the {\sc RVFit} code \citep{iglesias_marzoa15}. 
We tried several fits by leaving free and fixing the orbital period to that derived 
from the transit analysis but both gave too high reduced $\chi^2$ values because the
RV uncertainties are probably underestimated. The zero point of 
the calibration estimated from the sky spectrum of the template star has an error 
of $\sim$1.6 km/s. Therefore, to take this systematic uncertainty into account, 
we multiplied the RV errors given in Table \ref{tab_K2transitM5:ccfrv} by a factor of 10\@.
Those errors are of the order of or larger than potential RV variations induced
by the presence of spots\footnote{Zero order approximations show that the effect of star 
spots is approximately v$\sin$i$\times$dF, where dF is the difference in flux between the 
unspotted and spotted stellar surface. Estimates of dF ($<$3\%) and the rotation period 
($\sim$5 days) from our activity-conserved light curve and assuming no inclination translates 
into uncertainties of $<$0.12 km/s in the RV\@. Similarly, the eclipse depth could be biased 
by at most dF, affecting the radius estimate by $\sim$1.5\%.}
that can affect young M dwarfs, such as USco members.
In addition, we decided to discard the RV point obtained on 12 July 2015
with a phase close to 0.5 because the rms of the fit improves significantly.
The final reduced $\chi_\nu^2$ values are 1.5--1.8, being $\nu$ the
degrees of freedom, in this case, equal to 14\@.

%
%
\begin{figure}
  \centering
  \includegraphics[width=\linewidth, angle=0]{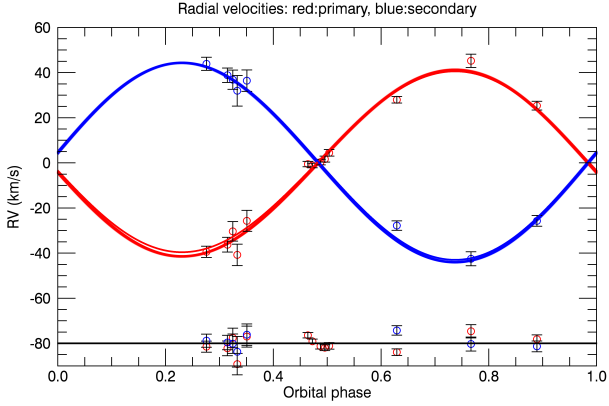}
   \caption{Radial velocity measurements from WHT/ISIS as a function of phase for the 
primary (blue symbols) and the secondary (red symbols). 
}
   \label{fig_K2transitM5:plot_Phase_vs_RV}
\end{figure}
\subsection{Physical parameters and global fits}
\label{K2transitM5:characterisation_parameters}

From the K2 light-curve, we infer that the radii and luminosities of both components are similar 
to less than a few percent. From the RV
measurements, we derive a maximum semiamplitude of 44.2$\pm$1.2 km/s and 40.8$\pm$1.4 km/s 
for the primary and secondary, respectively. Hence, the ratio of the maximum amplitude is 
0.92$\pm$0.03, corresponding to the mass ratio (q) of the system. This mass ratio is high 
and typical of low-mass and brown dwarf binaries \citep[see review by][and references therein]{burgasser07a}.

We used the K2 light curve corrected for the out of eclipse modulation (see
Sect.\ \ref{K2transitM5:K2_lc_phase}), the two photometric sequences obtained at the IAC80, 
and the RV data set to perform a combined fit and provide the final ephemeris and
physical parameters of the system. For the photometric eclipses, we modified the 
analytic formalisms of both \citet*{mandel02} and \citet{gimenez06a}, which are generally used 
to model transiting exoplanets, to include the effect of a luminous transiting object. For our 
purposes, we assign the name "primary" eclipses to the deepest ones, that happen when 
the hottest star is being eclipsed by the less luminous. We included a luminosity ratio 
$L_{21}$ of both stars (luminosity$_{\rm sec}$/luminosity$_{\rm pri}$), leading to the 
following corrected expressions

\begin{equation}
Model_{pri} = M + (1-M) \times L_{21} /(1+L_{21})
\end{equation}
 for the primary eclipses, and
\begin{equation}
Model_{sec} = M + (1-M)/(1+L_{21})
\end{equation}
for the secondary eclipses.

The final $\chi^{2}$ is evaluated as the sum of the $\chi^{2}$ of the individual data sets. For the 
K2 data set, and to avoid the blurring effect caused by the long integration times compared 
to the eclipse timescales \citep{kipping10a}, we oversampled the time vector to evaluate the 
model at five different points inside each time stamp, and then compute the average model value.

Using the \citet{gimenez06a} formalism, the parameters entering the fit were the orbital period, 
the epoch of primary eclipse, the phase of primary eclipse ingress $\phi_1$,
the orbital inclination, the eccentricity and argument of periastron, the $L_{21}$, the two 
masses of the stars, the time of periastron passage and an offset in RV\@. Two 
combinations of the quadratic limb darkening coefficients ($u_+ = u_a+u_b$ and $u_-=u_a-u_b$) 
were kept fixed to values extracted from the tables of \citet{claret12a}\footnote{We used fixed 
values of $u_+$\,=\,0.7 and $u_-$\,=\,0.15} . We used the {\tt{amoeba}}
downhill minimisation method \citep{csizmadia11} to find a minimum of the parameter space. 
The errors were estimated by several repetitions of the {\tt{amoeba}} minimisation, allowing 
the algorithm to start from slightly different initial positions to avoid the effect of falling into 
local minima. A total of 150 repetitions were performed, and we recorded the parameters 
and their $\chi^{2}$ at each step of the downhill algorithm. We selected a $\chi^{2}$ threshold 
as the maximum $\chi^{2}$ of all of the final convergence points of each of the {\tt{amoeba}}
trials. The uncertainties were finally estimated through gaussian fits to the distribution of 
each of the free parameters of the chains that produced a $\chi^{2}$ below this threshold. 
The revised transit duration is 2.4163 hours, given by the ``Phi'' parameters.
All the parameters for this formalism are quoted in Table \ref{tab_K2transitM5:physical_parameters_Gimenez2006}.

In the \citet{eastman13} re-implementation of the \citet*{mandel02} formalism, the parameters 
entering the fit were the log(period), the epoch of primary eclipse, the log of the system scale 
(a/r), the cosine of the orbital inclination, the $\sqrt(e)\cos w$ and $\sqrt(e)\sin w$, the $L_{12}$ 
ratio, the two masses of the stars, the time of periastron passage and an offset in RV\@.
The same quadratic limb darkening coefficients were also kept fixed. We used the EXOFAST implementations 
of the {\sc{amoeba}} minimisation algorithm to find a reasonable "global" 
minimum, and used this minimum as a starting point for a Differential Evolution Markov Chain 
method  \citep{braak06}. In this method, several Monte Carlo (MC) chains are run in parallel 
(twice the number of parameters entering the fit, in our case 22 chains), and they are used to 
evaluate the size of the steps on each MC iteration. The chains are evaluated for convergence 
when their number of independent draws is higher than 1000 and the Gelman-Rubin
statistic is less than 1.02 for all the parameters. After 60000 steps, the acceptance rate of 
the chains was 17\%, which is a reasonable value showing good mixing of the chains 
\citep{gelman03a}. To facilitate the comparison between the two methods, the uncertainties 
were estimated in the same way as in the first method, and the final values and their 
uncertainties are in Table \ref{tab_K2transitM5:physical_parameters_Mandel2002}. 
Most of the distributions are reasonably symmetrical so we only quote the mean 
uncertainties\footnote{The only exception being the argument of the periastron, which produces 
the following median and 34\% confidence intervals: 92.9$^{+1.3}_{-0.7}$}.
The values for the maximum amplitudes of the primary and secondary components agree within
1$\sigma$ with the values derived solely from the RV analysis.

%
%
\begin{table}
\begin{center}
\caption{Physical parameters with their error bars of the eclipsing system USco16m25 
derived using the formalism of \citet*{gimenez06a}.
}
\begin{tabular}{l c}
\hline
\hline
Parameters & Values and errors  \\
\hline
Period (d)              & 2.808862$\pm$0.000005 \cr
Epoch - 2454900         & 1997.52271$\pm$0.00014 \cr
K1  (km/s)              &    40.11$\pm$0.62 \cr
K2  (km/s)              &    43.64$\pm$1.03 \cr
Mass 1 (Solar mass)     &  0.091$\pm$0.005 \cr
Mass 2 (Solar mass)     &  0.084$\pm$0.004 \cr
Mass 1 (Jupiter mass)   &    95.6$\pm$5.5 \cr
Mass 2 (Jupiter mass)   &    87.6$\pm$4.3 \cr
q (mass ratio)          &   0.922$\pm$0.015 \cr
Omega (degs)            &    105.7$\pm$3.3 \cr
Eccentricity            &  0.011$\pm$0.002 \cr
Time periastron         &    1997.60$\pm$0.03 \cr
Orbital incl. (degs)    &    83.11$\pm$0.03 \cr
Gamma RV (km/s)         &   0.6$\pm$1.0 \cr
Surf.\ brightness ratio &    1.010$\pm$0.021 \cr
Luminosity ratio        &    0.966$\pm$0.016 \cr
Phi 1 (Orb. phi. 0-1)   &  0.01791$\pm$0.00006 \cr
Radius 1 / a            &  0.0828$\pm$0.0007 \cr
Radius 2 / a            &  0.0810$\pm$0.0007 \cr
Semimajor axis (A.U.)   &  0.0218$\pm$0.0004 \cr
Radius 1 (Solar radii)  &   0.388$\pm$0.008 \cr
Radius 2 (Solar radii)  &   0.380$\pm$0.008 \cr
Radius 1 (Jup. radii)   &   3.776$\pm$0.080 \cr
Radius 2 (Jup. radii)   &   3.695$\pm$0.080 \cr
Radius ratio            &   0.979$\pm$0.013 \cr
Density 1 (g cm-3)      &    2.20$\pm$0.06 \cr
Density 2 (g cm-3)      &    2.16$\pm$0.06 \cr
Impact parameter 1      &    1.456$\pm$0.010 \cr
Impact parameter 2      &    1.446$\pm$0.010 \cr
$\log$ g1 (cgs)         &    4.22$\pm$0.01 \cr
$\log$ g2 (cgs)         &    4.20$\pm$0.01 \cr
 \hline
 \label{tab_K2transitM5:physical_parameters_Gimenez2006}
 \end{tabular}
 \end{center}
\end{table}

%
%
%
\begin{table}
\begin{center}
\caption{Physical parameters with their error bars of the eclipsing system USco16m25
derived using the formalism of \citet*{mandel02}.
}
\begin{tabular}{l c}
\hline
\hline
Parameters & Values and errors  \\
\hline
Period (d)              & 2.808860$\pm$0.000007 \cr
Epoch - 2454900         & 1997.5227$\pm$0.0003 \cr
K1  (km/s)              &    40.11$\pm$1.28 \cr
K2  (km/s)              &    43.51$\pm$1.42 \cr
Mass 1 (Solar mass)     &  0.090$\pm$0.007 \cr
Mass 2 (Solar mass)     &  0.083$\pm$0.006 \cr
Mass 1 (Jupiter mass)   &     94.1$\pm$7.1 \cr
Mass 2 (Jupiter mass)   &    86.7$\pm$6.4 \cr
q (Mass ratio)          &   0.922$\pm$0.039 \cr
$\sqrt{e}$ * $\cos$ w   &  $-$0.013$\pm$0.002 \cr
$\sqrt{e}$ * $\sin$ w   &   0.236$\pm$0.038 \cr
Omega (degs)            &    93.1$\pm$0.9 \cr
Eccentricity            &  0.055$\pm$0.018 \cr
Time periastron         &    1997.52$\pm$0.014 \cr
Orbital incl.\ (degs)   &    83.2$\pm$0.1 \cr
a / R1                  &    12.26$\pm$0.11 \cr
Gamma RV (km/s)         &    1.4$\pm$0.6 \cr
Surf.\ brightness ratio &    1.48$\pm$0.23 \cr
Luminosity ratio        &   0.966$\pm$0.016 \cr
Radius 1 / a            &  0.0815$\pm$0.0007 \cr
Semimajor axis (A.U.)   &  0.0217$\pm$0.0005 \cr
Radius 1 (Solar radii)  &   0.381$\pm$0.009 \cr
Radius 2 (Solar radii)  &   0.312$\pm$0.027 \cr
Radius 1 (Jup. radii)   &   3.705$\pm$0.091 \cr
Radius 2 (Jup. radii)   &   3.038$\pm$0.265 \cr
Radius ratio            &   0.821$\pm$0.069 \cr
Density 1 (g cm-3)      &    2.30$\pm$0.07 \cr
Density 2 (g cm-3)      &    3.75$\pm$0.91 \cr
Impact parameter 1      &    1.455$\pm$0.011 \cr
Impact parameter 2      &    1.446$\pm$0.011 \cr
$\log$ g1 (cgs)         &    4.23$\pm$0.02 \cr
$\log$ g2 (cgs)         &    4.37$\pm$0.07 \cr
 \hline
 \label{tab_K2transitM5:physical_parameters_Mandel2002}
 \end{tabular}
 \end{center}
\end{table}

Both results are in agreement, and show that the system is on a grazing configuration.
The most discordant parameters are the luminosity ratio, the omega, and the eccentricity. We believe 
this is due to the MC runs exploring a region of the parameter space in which a more eccentric orbit 
with a periastron line a few degrees from the line of sight produces a slightly best $\chi^2$ as the 
other solutions. In such a configuration, the primary and secondary eclipses occur at different 
relative distances of the stars, and thus the flux of the occulted star is reduced by the square root 
of the distance when it is further away. As the depths of primary and secondary eclipses are comparable, 
the parameter that is changed to produce reasonable fits is the luminosity ratio of both components. 
From the very similar spectral types and the mass ratio $q$ close to 0.92, we deduce that luminosity 
ratios as high as 1.5 implied by the solution in Table \ref{tab_K2transitM5:physical_parameters_Mandel2002} 
are less realistic. Thus, for these three parameters, we rely on the values listed in 
Table \ref{tab_K2transitM5:physical_parameters_Gimenez2006}. It is important to note that despite these 
slight discrepancies, the physical values of both stars are in perfect agreement with the two explored 
formalism and methodologies. Our results are also in agreement with the independent analysis
of the system by \citet{david15c}.

A third fit to the light-curve only was also performed with UTM/UFIT 
\citep[Universal Transit Modeller / Universal Fitter;][]{deeg14} and gave binary parameters
in close agreement to Table \ref{tab_K2transitM5:physical_parameters_Gimenez2006}.
We warn that we kept the limb darkening coefficients fixed in both methods, relying on the
stellar model tabulations, and this could have influences on the obtained parameters. Despite
our efforts to select adequate limb darkening coefficients, the masses of the stars are in a
parameter space where the stellar models are poorly tested, and significant differences might
arise \citep{csizmadia13}. Additionally, the photometric curve shows out of eclipse modulations
that we interpreted as due to inhomogeneities on the stellar surfaces, and the stars are few
Myr old, again pointing towards a skeptical interpretation of the adopted limb darkening coefficient
values. This would affect mostly the measurements of the stellar radii, and we caution about
this extra noise source which is very difficult to overcome.
We limit the third light fraction to less than 10\%, given the absence of a third light emitter 
in our datasets. Within the small range of third light from 0 to 10\%, there are only negligible 
dependence of the inclination or the relative radius, (R1$+$R2)/a with third light. The values in 
Tables \ref{tab_K2transitM5:physical_parameters_Gimenez2006} and 
\ref{tab_K2transitM5:physical_parameters_Mandel2002} are given for the case of no third light. 
If any third light  would be present, we estimate that the relative radius would decrease by 0.05 
per 10\% of third light, whereas the inclination would increase by 0.5 degrees.

\subsection{USco16m25 in the realm of eclipsing binaries}
\label{K2transitM5:characterisation_discussion}

\begin{figure}[!h]
  \centering
  \includegraphics[width=\linewidth, angle=0]{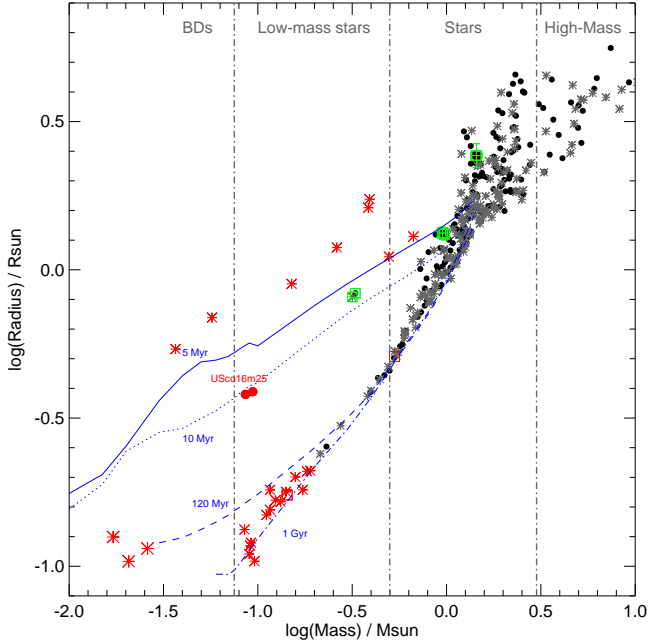}
   \caption{Mass-radius diagram for detached eclipsing binaries: filled dots
are primaries and asterisks are secondaries of systems with parameters determined
with an accuracy better than 2\%. Overplotted as red asterisks are field and young
low-mass eclipsing binaries as well as four known eclipsing brown dwarfs:
Corot-3\,b \citep[22 M$_{\rm Jup}$;][]{deleuil08},
KELT-1\,b \citep[27.38 M$_{\rm Jup}$;][]{siverd12}, and
Kepler-39\,b \citep[18 M$_{\rm Jup}$;][]{bouchy11b}.
Added in green are the only two eclipsing systems
known to date in the USco association. Overplotted in red are both components
of USco16m25: the size of the symbols indicates the size of the error bars.
Overplotted in blue are the solar-metallicity evolutionary tracks of the BT-Settl models
\citep{baraffe02} for two possible ages of 5 Myr (thick line) and 10 Myr (dashed line)
for USco, the age of the Pleiades (120 Myr; long-dashed line), and the age of field dwarfs
(1 Gyr; dot-dashed line).
}
   \label{fig_K2transitM5:plot_MRdiagram}
\end{figure}

In Fig.\ \ref{fig_K2transitM5:plot_MRdiagram} we plot a mass-radius diagram to place USco16m25
in context with other known eclipsing binaries \citep{southworth14a}\footnote{List of detached
eclipsing binaries taken from the census updated by John Southworth at
http://www.astro.keele.ac.uk/jkt/debcat/}. The filled dots and grey asterisks in
Fig.\ \ref{fig_K2transitM5:plot_MRdiagram} represent the primaries and secondaries of detached
eclipsing binaries with masses below 10 M$_{\odot}$ where most components have mass and radius
measurements accurate to 2\%. At lower masses, several eclipsing binaries do exist but the
uncertainties on their masses and radii vary and is usually greater than 2\%. Nonetheless, we
add several known systems to populate this part of the mass-radius diagram. We overplotted three 
transiting brown dwarfs with accurate masses (below 30 Jupiter masses) and radii 
\citep[Corot-3\,b; KELT-1\,b; Kepler-39\,b;][]{deleuil08,siverd12,bouchy11b}.
We also overplotted very-low mass components of the eclipsing binaries listed in Table 3 of
\citet{tal-or13} whose masses range from 0.08 to 0.2 M$_{\odot}$ 
\citep{pont05a,pont05b,hebb06,pont06a,maxted07,beatty07a,pyrzas09,pyrzas12a,parsons12a,parsons12b,parsons12c,ofir12a,triaud13a}
and added the M dwarf system from the WFCAM Transit Survey published by \citet{nefs13}.
We can see that most low-mass and brown dwarf components of eclipsing systems plotted in
Fig.\ \ref{fig_K2transitM5:plot_MRdiagram} follow the BT-Settl evolutionary models \citep{allard12} 
with ages between the age of the Pleiades \citep[120 Myr;][]{stauffer98,barrado04b} and the 
age of field stars ($\sim$1 Gyr).

%
%

We also added as red asterisks in Fig.\ \ref{fig_K2transitM5:plot_MRdiagram} three known young
eclipsing binaries \citep{king00a,irwin07a,cargile08,daRio09a,watson09a,gomez_maqueo12,gillen14}
with ages in the 1--20 Myr range \citep[see Table 1 in][]{stassun14a} along with the sole brown
dwarf eclipsing binary discovered to date \citep[2MASS\,J05352184$-$0546085;][]{stassun06,stassun07b}.
There are two known eclipsing systems recently discovered in the USco association
(green symbols in Fig.\ \ref{fig_K2transitM5:plot_MRdiagram}).
On the one hand UScoCTIO-5, a M4.5\,$+$\,M4.5 eclipsing binary with a period of 34 days and
masses of 0.3287$\pm$0.0024 and 0.3165$\pm$0.0016 M$_{\odot}$ and radii of 0.834$\pm$0.006
and 0.810$\pm$0.006 R$_{\odot}$ for the primary and secondary, respectively \citep{kraus15a}.
On the other hand, a triple system composed of two solar-type stars (masses of 0.950 and
0.975 M$_{\odot}$ and radii of 1.32--1.33 R$_{\odot}$) and a late-F star with a mass of
1.44 M$_{\odot}$ and a radius of 2.43 R$_{\odot}$ \citep{torres06,chen11a,alonso15a}
USco16m25 bridges the gap between the coolest low-mass eclipsing binaries younger than
10 Myr and the only brown dwarf eclipsing system known to date.

\begin{figure*}
  \centering
  \includegraphics[width=0.49\linewidth, angle=0]{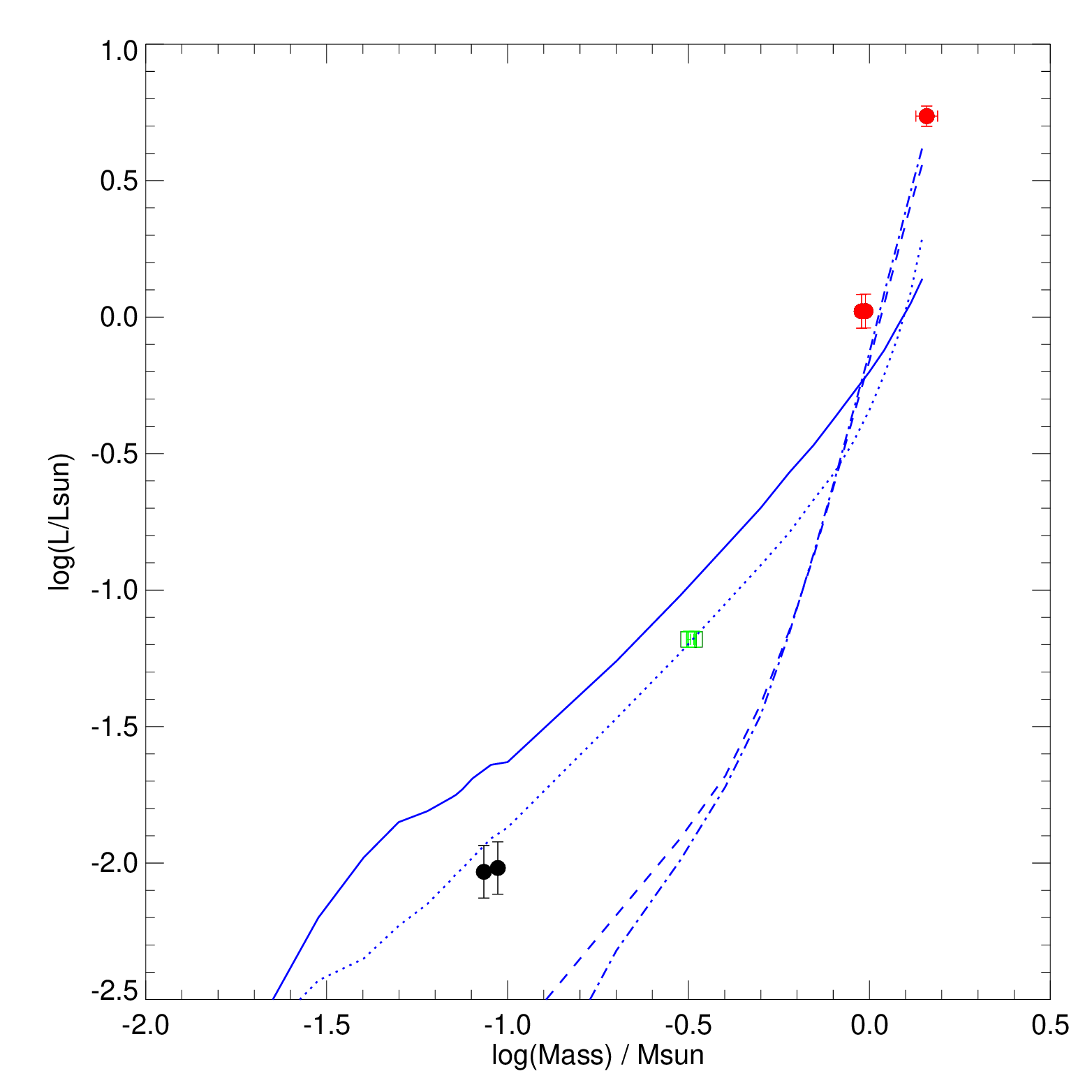}
  \includegraphics[width=0.49\linewidth, angle=0]{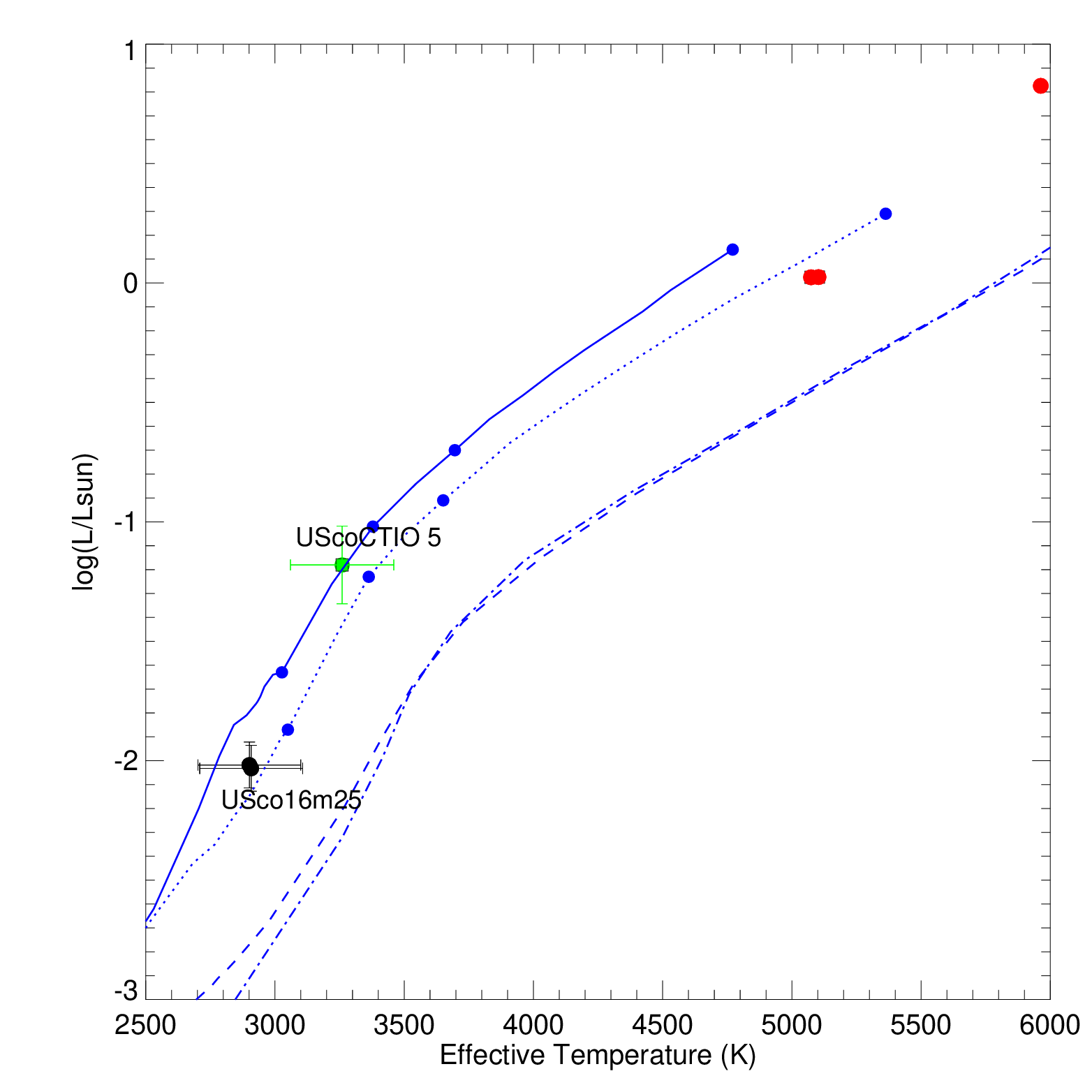}
   \caption{{\it{Left:}}
Mass-luminosity diagram for the three eclipsing binaries identified in USco by K2: USco16m25 is
represented as a black dot, UScoCTIO\,5 as a green square, and HD\,144588 as a red dot.
Overplotted in blue are the solar-metallicity evolutionary tracks of the BT-Settl models
\citep{baraffe02} for two possible ages of 5 Myr (thick line) and 10 Myr (dashed line)
for USco, the age of the Pleiades (120 Myr; long-dashed line), and the age of field dwarfs
(1 Gyr; dot-dashed line).
{\it{Right:}} Effective temperature vs luminosity diagram for the three USco systems
with the same colour set as in the left panel.
The blue dots on the 5 and 10 Myr isochrones correspond to masses of 1.4, 0.5, 0.3, 
and 0.1 M$_{\odot}$, respectively.
The size of the dots for the highest mass system represents the error bars on the measurements.
}
   \label{fig_K2transitM5:plot_MLTeff_diagram}
\end{figure*}

We derived the bolometric luminosity of the system using the $J$-band magnitude from UKIDSS,
assuming a distance of 145 pc for USco \citep{deBruijne97} and the most recent
bolometric correction for M dwarfs \citep[1.97$\pm$0.04 from the $r-J$ colour of USco16m25;][]{mann15a}.
We calculated a total luminosity of $\log$(L/L$_{\odot}$)\,=\,$-$1.724$\pm$0.096 dex,
yielding individual luminosities of $-$2.020$_{-0.121}^{+0.099}$ dex and $-$2.032$_{-0.121}^{+0.099}$ for the primary and secondary,
respectively. Using the Stefan-Boltzmann law, we derived an effective temperature of 
2901$_{-172}^{+199}$\,K and 2908$_{-172}^{+199}$\,K for the primary and secondary, respectively.
We plot USco16m25 in a mass-luminosity diagram on the left-hand side plot
in Fig.\ \ref{fig_K2transitM5:plot_MLTeff_diagram} along with the other two
eclipsing systems in USco (HD\,144588 and UScoCTIO\,5).

%
%
%

\begin{figure}
  \centering
  \includegraphics[width=\linewidth, angle=0]{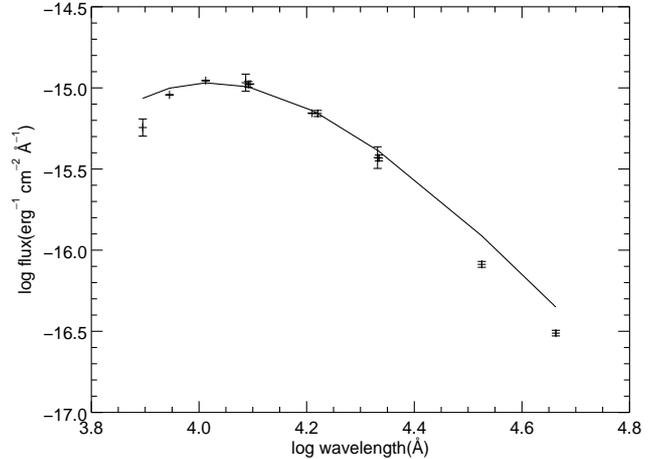}
   \caption{Spectral energy distribution of USco16m25\@. We obtained the fluxes and their associated
errors from the Spanish virtual observatory and calibrated them as incident fluxes on the Earth.
The wavelength ranges from the $I$-band (7862\AA{}) from the DENIS survey to the WISE $w2$-band
(4.6 microns). The solid line is the black-body fit.
}
   \label{fig_K2transitM5:plot_SED}
\end{figure}

We obtained an independent determination of the system's effective temperature (T$_{eff}$) and distance
from a fit of the flux from a spherical blackbody of given size and distance against the system's
spectral energy distribution (Fig.\ \ref{fig_K2transitM5:plot_SED}), obtained from Spanish Virtual
Observatory Sed Analyzer \citep[VOSA;][]{bayo08} over the 0.8--4.6 micron wavelength range
with data from the Deep Near-Infrared Southern Sky Survey \citep[DENIS;][]{DENISconsortium2005,epchtein97},
UKIDSS \citep{lawrence07}, and the Wide Field Infrared Survey \citep[WISE;][]{wright10}.
We note that this system is undetected in the reddest mid-infrared bands of WISE, region related to
the presence of a hot accretion disc. This fit to the spectral energy distribution  had two free parameters:
$T_{eff}$ and the ratio $R_{*} / d$, were $d$ is the distance. The fit indicated a temperature of 2800$\pm$100\,K
and a distance ratio of 6.5$\pm$0.5$\times$10$^{-11}$. Given the component sizes of 0.388$\pm$R$_{sol}$
listed in Table \ref{tab_K2transitM5:physical_parameters_Gimenez2006}, we inferred a distance of
135$\pm$10 pc. The distance estimate is strongly correlated with the temperature, in the sense
that higher temperatures imply further distances. The best fit from the BT-Settl models suggests
a temperature of 2800\,K, a gravity of $\log$\,g of 4.0, and a metallicity of $-$0.5 dex.
All these temperature estimates agree within the error bars. By comparison, the same method 
leads to a temperature of 3260\,K for UScoCTIO\,5 \citep{kraus15a}. Moreover, the triple
system HD\,144548 is composed of three components with $T_{eff}$ of 5943\,K, 5103\,K, and
5081\,K, respectively, using the $J$-band and bolometric corrections from \citet{pecaut13} to 
derive the bolometric luminosities. We compare the location of
the three eclipsing systems identified in USco by K2 in a temperature vs luminosity diagram on 
the right-hand side plot in Fig.\ \ref{fig_K2transitM5:plot_MLTeff_diagram}.

We note that UScoCTIO\,5 and USco16m25 lie close to the 5 and 10 Myr isochrones from 
the latest models of \citet{baraffe15} while HD\,144548 closer to older isochrones
(see Fig.\ \ref{fig_K2transitM5:plot_MLTeff_diagram}).
Although it is too early to extrapolate the age of the association based on the two coolest systems,
this estimate of 5--10 Myr is in line with the recent revision of the age of the USco association proposed
by \citet[11$\pm$2 Myr;][]{pecaut12} and the ages derived from isochrone fitting
\citep[5$\pm$2 Myr;][]{preibisch99,slesnick08,lodieu11a}. It is also consistent with the presence of
lithium in absorption in both components of USco16m25 and UScoCTIO\,5
(Sect.\ \ref{K2transitM5:characterisation_USco_memb}).

Both components of USco16m25 have masses about 3.5 times lower than UScoCTIO-5
and radii twice lower. The difference is fairly large considering the similar spectral types
of the two systems (M4--M4.5 vs M5.25), suggesting that the mass-radius relation decrease swiftly
close to the hydrogen-burning limit. USco16m25 is the lowest mass eclipsing binary in USco
and fills an important gap in the mass-radius in the 0.3--0.03 M$_{\odot}$ interval,
region occupied by three old objects
\citep[CM Draconis (M4.5\,$+$\,M4.5 and LSPM\,J1112$+$7626 ($>$\,M4);][]{morales09,terrien12a,irwin11}
and one secondary \citep[LSPM\,J1112$+$7626 ($>$\,M4);][]{irwin11} and two young systems
\citep[JW\,380 and 2MASS\,J05352184$-$0546085;][]{irwin07a,daRio09a,stassun06,stassun07b}.
Overall, USco16m25 represents a key eclipsing binary because it is the lowest mass member of
USco, filling the gap between the lowest mass stars and brown dwarfs in the mass-radius diagram.

We can conclude that the stellar/substellar boundary in USco lies at a spectral type of M5.5
or later, in agreement with the suggestion of \citet{luhman99a} purely based on theoretical
models. We can now compare this with the Pleiades \citep[125 Myr;][]{stauffer98} and
Orion (1-3 Myr), where eclipsing binaries
straddling the hydrogen-burning limit are known. In the former, PPl15 is a M6.5 spectroscopic
binary showing lithium in absorption at 670.8 nm \citep{stauffer94a,basri96,basri99b},
implying Pleiades brown dwarfs have spectral types equal or later than M7 \citep{stauffer98}.
In the latter, \citet{stassun06} discovered a M6.5 eclipsing binary in Orion,
suggesting that the stellar/substellar boundary lies at slightly earlier spectral type.
In conclusion, we invoke that the evolution of objects at the hydrogen-burning mass limit
evolve at very similar effective temperature in a narrow range in spectral type (M5.5--M7)
during the first 100Myr, something predicted by models \citep[e.g.][]{burrows93,chabrier97}.

%
%
%
%

%
%
\section{Conclusions and future work}
\label{K2transitM5:conclusions}

We have exploited the K2 light curves from campaign two and discovered an eclipsing binary
made of two very-low mass star members of the young USco association that we confirmed
with ground-based RV follow-up. The main results of our work can be summarised as follows:
\begin{enumerate}
\item [$\bullet$] USco16m25 is the coolest eclipsing binary in USco with a combined spectral
type of M5.25$\pm$0.25
\item [$\bullet$] we inferred a period of $\sim$2.81 days and transit duration of $\sim$2.4h from the analysis 
of the K2 light curve
\item [$\bullet$] we estimate comparable radii, luminosities, and effective temperatures to within 
5\% for both components 
of the binary
\item [$\bullet$] we derived masses of 0.091$\pm$0.005 M$_{\odot}$ and 0.084$\pm$0.004 M$_{\odot}$
for the primary and secondary component, yielding a mass ratio $q$ of 0.922$\pm$0.015
\item [$\bullet$] objects straddling the stellar/substellar boundary evolve at very similar 
effective temperature in a narrow range in spectral type during the first 100Myr
\item [$\bullet$] USco16m25 bridges the gap between the coolest low-mass eclipsing binaries
and the only brown dwarf eclipsing system known at ages younger than 10 Myr
\item [$\bullet$] USco16m25 is the third eclipsing system in USco with masses and radii obtained
independently of models, suggesting that members of the USco association share an age around 5--10 Myr
\end{enumerate}

We plan to increase the number and accuracy of our RV curves by obtaining optical
spectra at higher resolution during a longer period of time covering several days to refine the
the physical parameters of the USco16m25 system. In the course of these RV measurements, 
we also plan to measure the spin-orbit (mis)-alignment of USco16m25 by modeling the change in 
shape of spectral lines \citep[see e.g.][]{albrecht14}. Such a measurement for a young eclipsing 
binary system will provide us an important insight on the initial spin-orbit (mis)-alignment of 
multiple systems at the stellar/substellar boundary.
We also aim at extending our analysis of the Kepler light curves to other clusters at
different ages, e.g.\ the Pleiades \citep[125$\pm$8 Myr;][]{stauffer98}, Praesepe 
\citep[590--800 Myr;][]{fossati08,brandt15a}, and the Hyades \citep[625$\pm$50 Myr;][]{maeder81},
which are included in the fourth and fifth K2 campaigns to populate the mass-radius diagram at
low masses and at different ages.

%
%
\begin{acknowledgements}
NL was funded by the Ram\'on y Cajal fellowship number 08-303-01-02\@.
J.\ I.\ G.\ H.\ acknowledges financial support from the Spanish Ministry of Economy and 
Competitiveness (MINECO) under the 2011 Severo Ochoa Program MINECO SEV-2011-0187\@.
H.\ D.\ acknowledges support from MINECO under grant AYA2012-39346-C02-02\@.
N.\ N.\ acknowledges supports by the NAOJ Fellowship, Inoue Science Research Award, and 
Grant-in-Aid for Scientific Research (A) (JSPS KAKENHI Grant Number 25247026).
This work was performed, in part, under contract with the Jet Propulsion Laboratory (JPL) 
funded by NASA through the Sagan Fellowship Program executed by the NASA Exoplanet Science Institute.
Y.\ K.\ acknowledges supports by Grant-in-Aid for JSPS Fellows (JSPS KAKENHI Grant Number 15J08463) 
and the Leading Graduate Course for Frontiers of Mathematical Sciences and Physics.
NL thanks Kevin Luhman for kindly providing his spectra of USco members prior to publication.

This paper includes data collected by the Kepler mission. Funding for the Kepler mission is 
provided by the NASA Science Mission directorate. The authors gratefully acknowledge the 
team who made Kepler possible, including the extension of the mission and the K2 campaign.

This work is based on observations made with the Gran Telescopio Canarias (GTC), 
operated on the island of La Palma in the Spanish Observatorio del Roque de los 
Muchachos of the Instituto de Astrof\'isica de Canarias (program GTC38-15A; PI Lodieu).

The WHT and its service programme are operated on the island of La Palma by the Isaac 
Newton Group in the Spanish Observatorio del Roque de los Muchachos of the Instituto 
de Astrof\'isica de Canarias (service program SW2015a31 and DDT program).  
We warmly thank the WHT staff for carrying out our observations.

This article is based on observations made with the IAC80 and Carlos Sanchez telescopes 
operated on the island of Tenerife by the Instituto de Astrof\'isica de Canarias in the Spanish 
Observatorio del Teide. 

This research has been carried out using telescope time awarded by the CCI
International Time Programme at the Canary Islands Observatories (program ITP13-8;
PI: Rubi\~no-Martin). The WHT observer on 17 June 2015 was Antonio Ferragamo.
We also thank Rafael Barrena for his help.


This publication makes use of VOSA, developed under the Spanish Virtual Observatory project 
supported from the Spanish MICINN through grant AyA2011-24052\@.
This research has made use of the Simbad and Vizier databases, operated
at the Centre de Donn\'ees Astronomiques de Strasbourg (CDS), and
of NASA's Astrophysics Data System Bibliographic Services (ADS).
\end{acknowledgements}
%

%
%
\bibliographystyle{aa}
\bibliography{../../AA/mnemonic,../../AA/biblio_old}
%

\end{document}